\documentclass[%
preprint,
 longbibliography,
 amsmath,amssymb,
 10pt,
]{revtex4-1}

\usepackage{graphicx}
\usepackage{dcolumn}
\usepackage{bm}
\usepackage{bbold}
\usepackage{braket}
\usepackage{mathtools}
\usepackage{setspace}
\usepackage[colorinlistoftodos]{todonotes}

\begin{document}

\preprint{APS/123-QED}

\title{Photon blockade in weakly-driven cavity QED systems with many emitters}

\author{Rahul Trivedi}\email{rtrivedi@stanford.edu}
\affiliation{E. L. Ginzton Laboratory, Stanford University, Stanford, CA 94305, USA}
\author{Marina Radulaski}
\affiliation{E. L. Ginzton Laboratory, Stanford University, Stanford, CA 94305, USA}
\author{Kevin A. Fischer}
\affiliation{E. L. Ginzton Laboratory, Stanford University, Stanford, CA 94305, USA}
\author{Shanhui Fan}
\affiliation{E. L. Ginzton Laboratory, Stanford University, Stanford, CA 94305, USA}
\author{Jelena Vu\v{c}kovi\'c}
\affiliation{E. L. Ginzton Laboratory, Stanford University, Stanford, CA 94305, USA}

\date{\today}

\begin{abstract}
We use the scattering matrix formalism to analyze photon blockade in coherently-driven CQED systems with a weak drive. By approximating the weak coherent drive by an input single- and two-photon Fock state, we reduce the computational complexity of the transmission and the two-photon correlation function from exponential to polynomial in the number of
emitters. This enables us to easily analyze cavity-based systems containing $\sim$50 quantum emitters with modest computational resources. Using this approach we study the coherence statistics of photon blockade while increasing the number of emitters for resonant and detuned multi-emitter CQED systems --- we find that increasing the number of emitters worsens photon blockade in resonant systems, and improves it in detuned systems. We also analyze the impact of inhomogeneous broadening in the emitter frequencies on the photon blockade through this system.
\end{abstract}

\maketitle


\emph{Introduction.~}Cavity quantum electrodynamics (CQED) is a fundamental model of light and matter which has been experimentally implemented in a variety of physical platforms. Atomic and solid state CQED systems with a few two-level emitters have exhibited a rich set of quantum phenomena in transmission statistics, including, but not limited to, the vacuum Rabi oscillations \cite{brune1996quantum, yoshie2004vacuum}, the conventional and the unconventional photon blockade \cite{birnbaum2005photon, imamoǧlu1998erratum, snijders2018observation}, and the photon-induced tunneling \cite{faraon2008coherent}. While suitable approximations can provide understanding of the eigenstructure of multi-element CQED systems \cite{diniz2011strongly, radulaski2017nonclassical} obtained in experiments \cite{thompson1992observation, zhong2017interfacing}, the numerical studies of light-emission and scattering from this system have been limited due to the exponential scaling of the Hilbert space with the number of emitters.

The scattering matrix formalism for quantum-optical systems provides the solution to this problem. Recently, a general formalism for computing this scattering matrix for an arbitrary time-independent and time-dependent Markovian quantum-optical system was developed \cite{xu2015input2, trivedi2018few}, reducing its computation to that of an effective propagator for the quantum-optical system. Use of the scattering matrices allows relating the transmission and two-photon correlation through a system to the single- and two-photon scattering matrix whose computation time scales as $\sim\mathcal{O}(N^3)$ and $\sim \mathcal{O}(N^6)$ respectively in the number of emitters $N$.

In this letter, we use the scattering matrix formalism to study multi-emitter CQED systems with a large number of emitters ($N\sim$ 50) driven by weak continuous-wave classical light (e.g. a laser). We show that increasing the number of emitters does not increase the depth of the photon blockade in a resonant multi-emitter CQED systems with identical emitters. However, we find that increasing the number of emitters improves photon blockade if the emitters are detuned from the cavity resonance. Finally, we study the impact of inhomogenous broadening \cite{evans2016narrow, zhang2018strongly, dory2018optimized, radulaski2017scalable, bracher2017selective, banks2018resonant, lukin20194h} in the emitter frequencies on photon blockade in the multi-emitter systems.

\begin{figure}[b]
    \centering
    \includegraphics[scale=0.55]{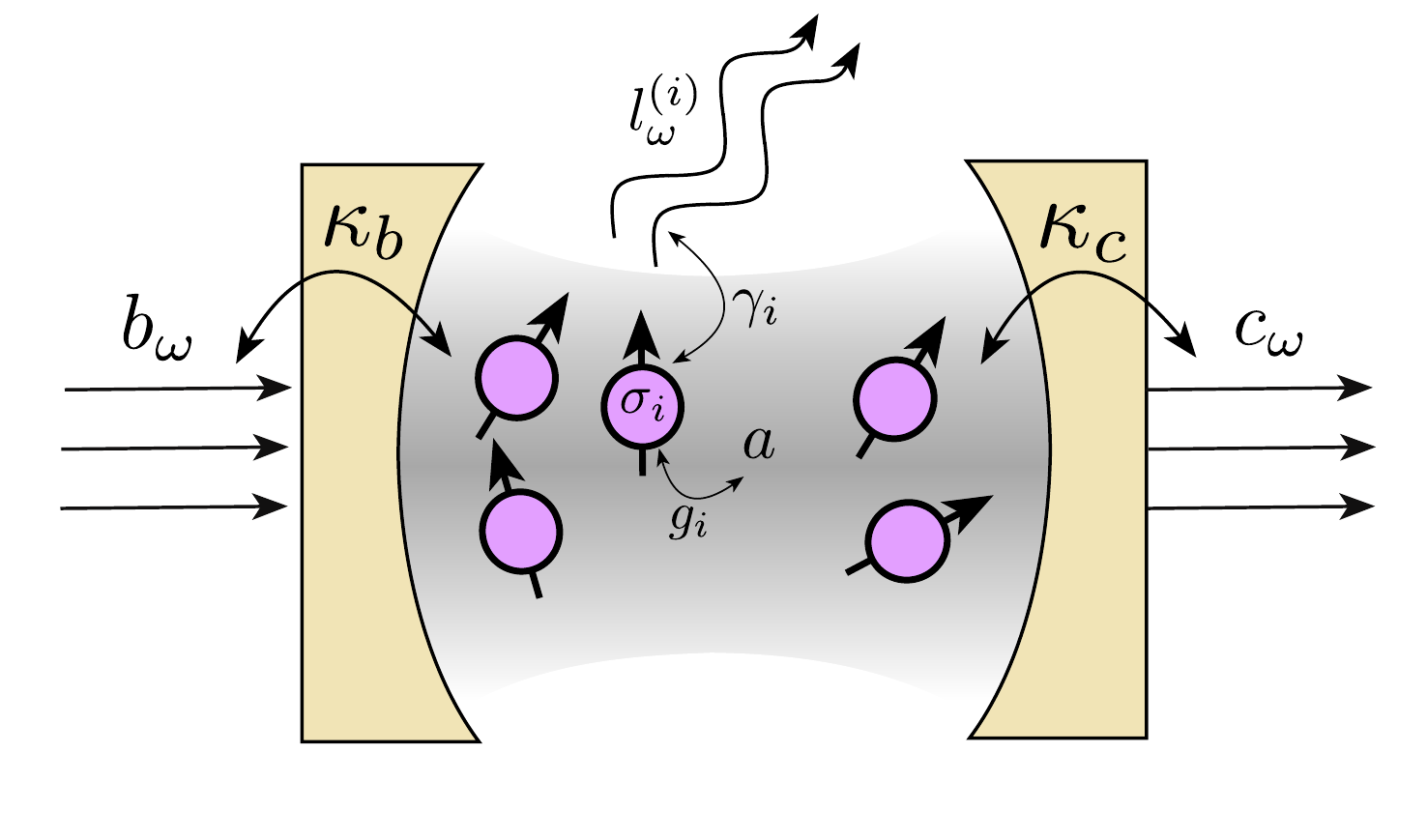}
    \caption{Schematic of the multi-emitter CQED system. An optical cavity mode couples to $N$ emitters with coupling constants $g_i$, $1 \le i \le N$. The input and output coupling constants are $\kappa_b$ and $\kappa_c$. In addition to the cavity, the emitters also couple to loss-channels with coupling constants $\gamma_i$. The total decay rate of the cavity is given by $\kappa = \kappa_b + \kappa_c$ (we assume $\kappa_b = \kappa_c = \kappa / 2$ throughout this paper).}
    \label{fig:schematic}
\end{figure}

\emph{Simulation method.} A schematic of the considered system is shown in Fig.~\ref{fig:schematic} --- a cavity, with annihilation operator $a$, is coupled to $N$ two-level emitters, with lowering operators $\sigma_i$, $1\leq i \leq N$. The cavity is excited through a waveguide, with a frequency dependent annihilation operator $b_\omega$, and the emission from the cavity is collected through another waveguide, with annihilation operator $c_\omega$. The emitters, in addition to coupling to the cavity mode, also radiate into loss channels with annihilation operators $l^{(i)}_\omega$ --- these loss channels model the linewidths of the emitters. The Hamiltonian for the multi-emitter CQED system is given by:
\begin{align}
H_\text{sys} = \omega_c a^\dagger a + \sum_{i=1}^N \bigg[\omega_i \sigma_i^\dagger \sigma_i + g_i \big(a \sigma_i^\dagger + \sigma_i a^\dagger \big) \bigg].
\end{align}
where $\omega_c$ is the cavity resonance frequency, $\omega_i$ is the transition frequency of the $i^\text{th}$ emitter and $g_i$ is the coupling constant between the $i^\text{th}$ emitter and the cavity mode. We study the excitation of this system with a continuous-wave coherent state at frequency $\omega_L$, described by an input state:
\begin{align}
\ket{\psi_\text{in}} = \exp[\beta_0 (b_{\omega_L}^\dagger - b_{\omega_L})] \ket{\text{vac}}
\end{align}
\begin{figure*}[htpb]
    \centering
    \includegraphics[scale=0.22]{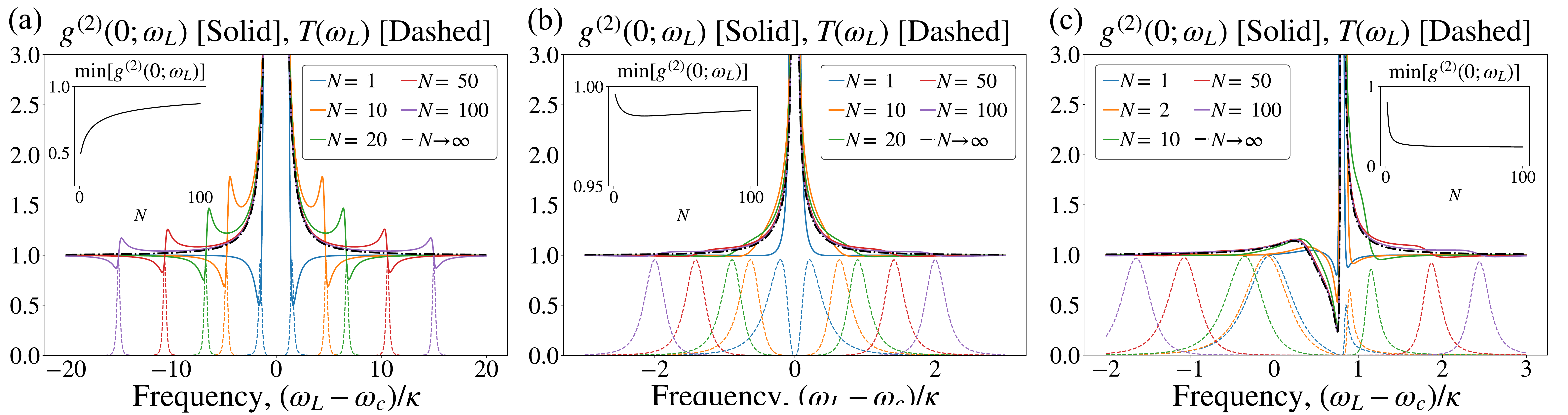}
    \caption{{Equal-time correlation $g^{(2)}(0; \omega_L)$ and transmissivity $T(\omega_L)$ for (a) strongly coupled resonant emitters ($g = 2 \kappa, \omega_e = \omega_c$), (b) weakly coupled resonant emitters ($g = 0.2 \kappa, \omega_e = \omega_c$) and (c) weakly coupled detuned emitters ($g = 0.2\kappa, \omega_e - \omega_c = 0.8\kappa$). The insets show the dependence of $\min_{\omega_L}[g^{(2)}(0; \omega_L)]$ as a function of $N$ and the dashed lines show $\lim_{N\to\infty}g^{(2)}(0; \omega_L)$ computed using Eq.~\ref{eq:g2_lim}. Increasing the number of emitters clearly deteriorates the polaritonic photon blockade observed in the system. For detuned systems, increasing the number of emitters enhances the interference based blockade. $\gamma = 0.01\kappa$ is assumed in all simulations.}}
    \label{fig:coupling_strength}
\end{figure*}
where $\beta_0^2$ is the photon flux (number of photons per unit time) in the coherent state. As is detailed in the supplementary material \cite{supp}, we establish the following relationship of the transmission $T(\omega_L)$ and two-photon correlation $g^{(2)}(t_1, t_2; \omega_L)$ to the single-photon $[S_{c}(\cdot)]$ and two-photon $[S_{c, c}(\cdot)]$ scattering matrices for continuous-wave input in the limit of small input photon flux $\beta_0^2$:
\begin{subequations}
\begin{align}
    &T(\omega_L) = \bigg|\int_{t'=-\infty}^\infty S_c(t; t') \exp(-\textrm{i}\omega_L t') \textrm{d}t' \bigg|^2 \\
    &g^{(2)}(t_1, t_2; \omega_L) = \frac{1}{4T^2(\omega_L)}  \times \nonumber\\
    &\bigg|\int_{t_1', t_2'=-\infty}^\infty S_{c,c}(t_1, t_2; t_1', t_2') \exp[-\textrm{i}\omega_L (t_1' + t_2')] \textrm{d}t_1' \textrm{d}t_2' \bigg|^2
\end{align}
\end{subequations}
where the $S$ matrices capture scattering of photons propagating in the input-waveguide (with annihilation operator $b_\omega$) to the output-waveguide (with annihilation operator $c_\omega$). The scattering matrices are functions only of the system operators and external coupling constants $\kappa_{b,c}$ and $\gamma_n$.

The dominant cost for computing these scattering matrices is that of diagonalizing the effective Hamiltonian $H_\text{eff}$ \cite{supp}:
\begin{align}
    H_\text{eff} = H_\text{sys} - \frac{\textrm{i}\kappa}{2} a^\dagger a - \sum_{i=1}^N \frac{\textrm{i}\gamma_i}{2} \sigma_i^\dagger \sigma_i
\end{align}
where $\kappa = \kappa_b + \kappa_c$ is the total decay rate for the optical cavity. Since $H_\text{eff}$ conserves the total excitation number ($a^\dagger a + \sum_{n=1}^N \sigma_n^\dagger \sigma_n)$, this diagonalization can be performed separately within the excitation conserving subspaces of the full Hilbert space. When computing the single- and two-photon scattering matrices, it is only necessary to diagonalize the effective Hamiltonian within the single- and two-excitation subspaces the cost of which approximately as $\sim \mathcal{O}(N^3)$ and $\sim\mathcal{O}(N^6)$ respectively. {We note that when the emitters are identical (i.e.~$\omega_i = \omega, \gamma_i = \gamma$ and $g_i = g$ \ for all $i \in \{1, 2 \dots N\}$), by utilizing the Clebsh-Gordan series this diagonalization can be mapped to the diagonalization of $3\times 3$ and $2\times 2$ complex matrices \cite{supp}.}

 {Having diagonalized $H_\text{eff}$, the transmission $T(\omega_L)$ and equal-time two-photon correlation $g^{(2)}(0; \omega_L) = g^{(2)}(t, t; \omega_L)$ can be expressed as \cite{supp}:
\begin{subequations}
\begin{align}
    &T(\omega_L) =  \kappa_b \kappa_c \bigg|\sum_{i = 1}^{\mathcal{N}_1}\frac{(\bra{\text{G}} a \ket{\phi^{(1)}_i}_\text{T})^2}{\lambda_i^{(1)}-\omega_L} \bigg|^2 \\
    &g^{(2)}(0; \omega_L) = \bigg|\sum_{i=1}^{\mathcal{N}_2}\Gamma_i(\omega_L) \bigg|^2\label{eq:g2}
\end{align}
\end{subequations}
where $\ket{\text{G}}$ is the ground state of the multi-emitter CQED system, $\langle\cdot \rangle_\text{T}$ denotes a `transpose' inner product between two states, $\mathcal{N}_i$ is the dimensionality of the $i^\text{th}$ excitation subspace of the multi-emitter CQED system,  $(\lambda^{(i)}_j, \ket{\phi^{(i)}_j})$ are the eigenvalues and eigenstates of $H_\text{eff}$ within the $i^\text{th}$ excitation subspace and $\Gamma_i(\omega_L)$, given below, can be interpreted as the contribution of the $i^\text{th}$ two-excitation eigenstate to the equal time two-photon emission:
\begin{align}\label{eq:gamma}
    \Gamma_i(\omega_L) =& \frac{{\kappa_b \kappa_c}}{T(\omega_L)}\Bigg(\frac{\bra{\text{G}}a^2\ket{\phi^{(2)}_i}_\text{T}}{\lambda^{(2)}_i-2\omega_L}\Bigg)\times \nonumber\\ &\sum_{j=1}^{\mathcal{N}_1}\Bigg(\frac{\bra{\phi^{(2)}_i}a^\dagger\ket{\phi^{(1)}_j}_\text{T}\bra{\phi^{(1)}_j}a^\dagger\ket{\text{G}}_\text{T}}{\lambda^{(1)}_j - \omega_L}\Bigg)
\end{align}
These expressions for $T(\omega_L)$ and $g^{(2)}(0; \omega_L)$ explicitly show their dependence on the energy eigenvalues [$\sim\text{Re}(\lambda_i^{(j)})$], linewidths [$\sim\text{Im}(\lambda_i^{(j)})$] as well as the eigenstates ($\ket{\phi_j^{(i)}}$) of the multi-emitter CQED systems.}

\begin{figure*}[htpb]
    \centering
    \includegraphics[scale=0.285]{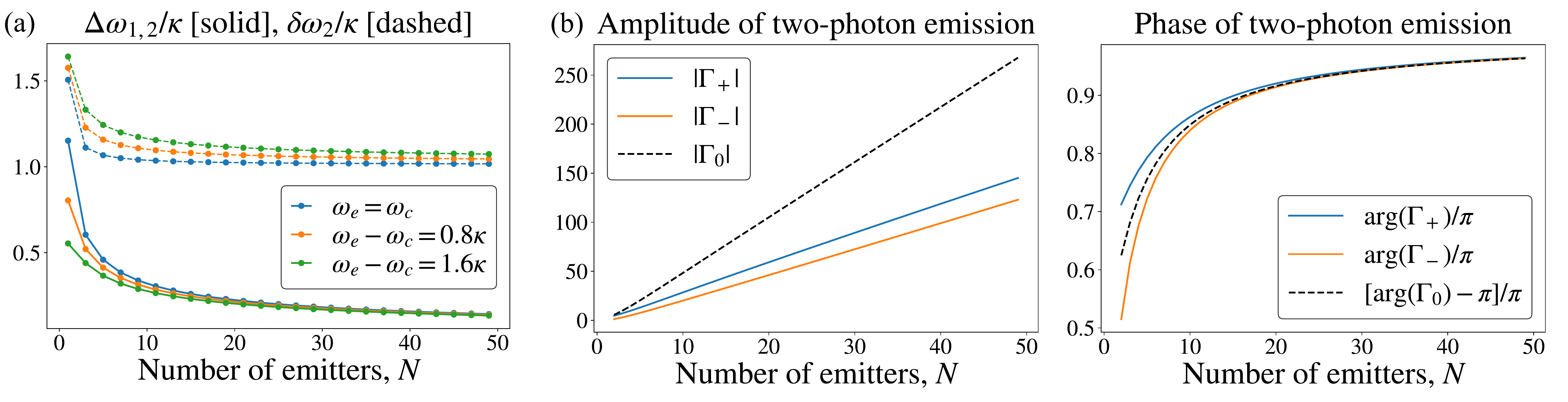}
    \caption{
    {(a) Anharmonicity $\Delta \omega_{1, 2}$ and the linewidth $\delta \omega_2$ of the most harmonic eigenstate in the second excitation subspace as a function of number of emitters. We have used $g = 2\kappa$ and only considered those two-excitation eigenstates which have a non-zero overlap with two photons in the cavity while computing $\Delta \omega_{1, 2}$ and $\delta \omega_2$. (b) Amplitude and phase of the equal-time two-photon emission ($\Gamma_\pm, \Gamma_0$) from the three eigenstates ($\ket{\phi_\pm^{(2)}}$, $\ket{\phi_0^{(2)}}$) that contribute to $g^{(2)}(0; \omega_L)$ in a detuned multi-emitter CQED system with weakly coupled emitters ($g = 0.2\kappa$, $\omega_e - \omega_c = 0.8\kappa$) at the frequency corresponding to the interference based blockade. $\gamma=0.01\kappa$ is assumed in all simulations.}}
    \label{fig:explanation}
\end{figure*}

\emph{Results}. Using a large number of identical emitters coupling coherently to the same cavity mode is a potential strategy to achieve strong coupling between the emitters and the cavity in a situation where an individual emitter only weakly couples to the cavity mode. {Figure~\ref{fig:coupling_strength}(a)-(b) shows the transmissivity $T(\omega_L)$ and equal-time correlation $g^{(2)}(0; \omega_L)$ for multi-emitter CQED systems with 1--100  emitters. Consistent with the result obtained on a direct diagonalization of $H_\text{sys}$, we observe that the splitting between the polaritonic peaks in the transmissivity scales as $\sqrt{N}$. We also observe that the minimum two-photon correlation $g^{(2)}(0; \omega_L)$, which is achieved at the polaritonic frequencies, tends towards unity with an increase in the number of emitters for both strongly-coupled emitters [Fig.~\ref{fig:coupling_strength}(a)] and weakly-coupled emitters [Fig.~\ref{fig:coupling_strength}(b)]. This trend can be easily explained by looking closely at the `anharmonicity' ($\delta \omega_{1, 2}$) between the single- and two-excitation eigenenergies of the multi-emitter CQED system:
\begin{align}
    \Delta \omega_{1, 2} = \min_{i, j} \big|\text{Re}[2\lambda^{(1)}_i] - \text{Re}[\lambda^{(2)}_j]\big|.
\end{align}
Figure~\ref{fig:explanation}(a) shows $\Delta \omega_{1, 2}$ as a function of $N$ along with the linewidth $\delta\omega_{2} = \text{Im}[2\lambda^{(2)}_i]$ of the most harmonic eigenstate in the two-excitation subspace --- increasing the number of emitters makes the system's energy levels more equally spaced while saturating their linewidths, thereby worsening photon blockade. It is worth noting that with the weakly coupled emitters the value of the min$[g^{(2)}(0; \omega_L)]$ has an initial decrease, before monotonically increasing with the number of emitters consistent with previously reported results \cite{radulaski2017nonclassical}. We also observe a pronounced `bunching' peak in $g^{(2)}(0; \omega_L)$ for strongly coupled emitters [Fig.~\ref{fig:coupling_strength}(a)], near the anti-bunching dip --- this corresponds to $2\omega_L$ being resonant with the two-excitation eigenstates. $g^{(2)}(0;\omega_L)$ at the bunching peak also tends to $1$ as $N\to\infty$ due to the system eigenstates becoming increasingly harmonic. Moreover, $g^{(2)}(0; \omega_L)$ can be analytically evaluated in the limit of $N\to \infty$ to obtain \cite{supp}:
\begin{align}\label{eq:g2_lim}
    \lim_{N \to \infty}g^{(2)}(0; \omega_L) =\bigg |1 - \frac{g^2}{(\omega_L - \lambda_e)(2\omega_L- \lambda_e-\lambda_c)} \bigg |^2
\end{align}
where $\lambda_e = \omega_e - \textrm{i}\gamma /2$ and $\lambda_c = \omega_c - \textrm{i}\kappa / 2$. As can be seen from Figs.~\ref{fig:coupling_strength}(a) and (b), in the limit of large number of emitters, the multi-emitter system does not show any blockade --- photon bunching can be seen at $\omega_L \sim \omega_c$ as a consequence of near zero single-photon transmission.
}

\begin{figure*}[htpb]
    \centering
    \includegraphics[scale=0.06, width=\textwidth]{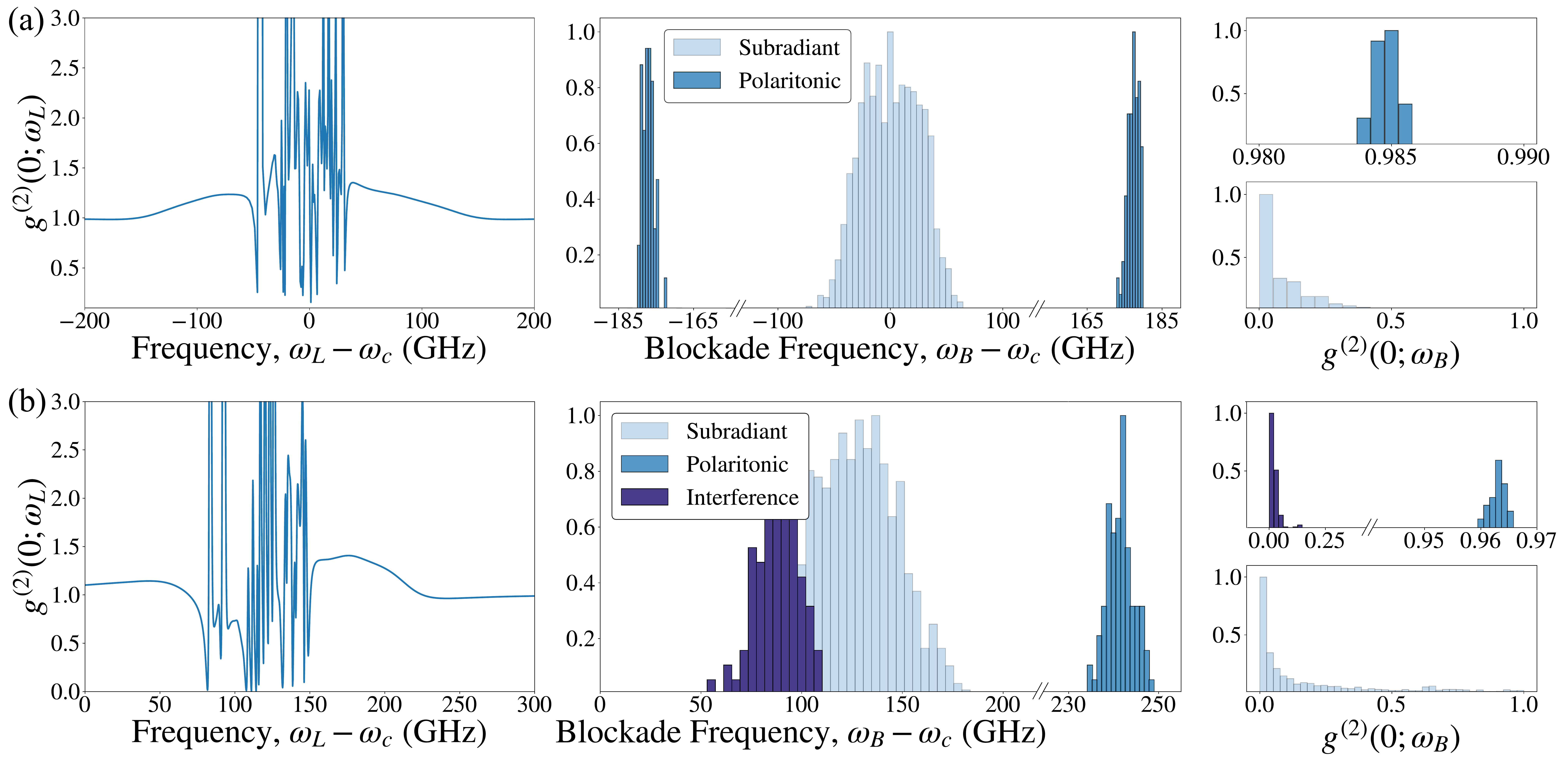}
    \caption{{Impact of inhomogeneous broadening on the photon blockade in multi-emitter CQED systems for (a) the emitters, on an average, being resonant with the cavity. (b) emitters that are, on an average, detuned from the cavity resonance by $\langle\omega_e\rangle - \omega_c = 0.8\kappa = 2\pi \cdot 20$ GHz. For both cases, we show a typical lineshape $g^{(2)}(0;\omega_L)$, and the statistics of the frequencies $\omega_B$ and the $g^{(2)}(0; \omega_B)$ values for the polaritonic and subradiant photon blockade (Note that the $y$-axis in the histogram is the unnormalized frequency of occurence of the sample statistic). Parameter values $\Delta = 25$ GHz, $\kappa = 2\pi \cdot 25$ GHz, $g = 0.2 \kappa = 2\pi \cdot 5$ GHz and $\gamma = 2\pi \cdot0.3$ GHz are assumed in all simulations.}}
    \label{fig:inhom_broadening}
\end{figure*}

We next study the impact of detuning between the emitters and the optical mode on the polaritonic photon blockade [Fig.~\ref{fig:coupling_strength}(c)]. Consider the transmission lineshape: there is a distinct Fano dip to nearly zero transmission at $\omega_L \approx \omega_e$ where the single-photon transmission through the two single-excitation eigenstates (one being more cavity-like and the other being more emitter-like) exactly cancels. Towards the right of this Fano dip, the light antibunches ($g^{(2)}(0; \omega_L) < 1$) owing to the standard polaritonic photon blockade of a detuned system \cite{laussy2012climbing,muller2015coherent}. Due to the multi-emitter system becoming more harmonic with an increase in the number of emitters [Fig.~\ref{fig:explanation}(a)], this blockade effect degrades similar to that in the resonant CQED system. {Photon bunching ($g^{(2)}(0; \omega_L) > 1$) is observed exactly at the Fano dip due to the single-photon transmission becoming nearly zero --- within the framework of scattering theory, this is equivalent to the contribution of the unconnected (linear or frequency-preserving) part of the scattering matrix being small in the output state, and the scattering happening almost entirely from the connected (nonlinear or frequency-mixing) part of the two-photon $S$ matrix  \cite{shen2007strongly}.

Slightly left of the Fano dip, we again observe photon antibunching --- moreover, unlike polaritonic blockade, the blockade depth increases with increasing $N$. This blockade occurs due to a destructive interference between the two-photon emissions from different two-excitation eigenstates. More insight into this phenomena can be obtained by closely studying the two-excitation eigenstates as well as their contribution to two-photon emission. Note from Eqs.~\ref{eq:g2} and \ref{eq:gamma} that only the two-excitation eigenstates which have non-zero overlap with two-photons in the cavity (i.e. $\bra{\text{G}}a^2 \ket{\phi_i^{(2)}} \neq 0$) have a non-zero contribution to $g^{(2)}(0)$. When all the emitters are identical, there are three such two-excitation eigenstates --- $\ket{\phi^{(2)}_\pm}$ which have a probability of $1/4$ of having 2 photons in the cavity in the limit of $N \to \infty$, and $\ket{\phi^{(2)}_0}$ which when $N \to \infty$ has a probability of $1/2$ of having 2 photons in the cavity \cite{supp}. Figure~\ref{fig:explanation}(b)-(c) shows the amplitude and phase of the contribution of these three eigenstates to equal-time two-photon emission (i.e.~$\Gamma_i(\omega_L)$ defined by Eq.~\ref{eq:gamma}) at the blockade frequency --- it can easily be seen that individually the eigenstates have significant two-photon emission, with the amplitude of emission being proportional to the probability of the eigenstate having 2 photons in the cavity mode. However, the two-photon emission from $\ket{\phi_0^{(2)}}$ is out of phase from the emission of $\ket{\phi_\pm^{(2)}}$, with the phase difference between them approaching $\pi$ as the number of emitters increases. This explains the interference based character of this blockade, as well as its dependence on $N$. The limit of $g^{(2)}(0; \omega_L)$ as $N \to \infty$, given by Eq.~\ref{eq:g2_lim} and plotted in Fig.~\ref{fig:coupling_strength}(c), also shows a pronounced interference based antibunching, along with a disappearance of the polaritonic blockade. Moreover, the interference based blockade can be made deeper by increasing the detuning between the emitters and the cavity mode \cite{supp}. We also note that the transmission $T(\omega_L)$ at the interference based photon blockade is small --- it scales in inverse proportion to $N^2$ \cite{supp} --- there thus exists a tradeoff between the purity of the single-photon state emitted by these systems and the brightness of the single-photon state.}

While the previous analysis was primarily done under the assumption of identical emitters, emitters in practical systems are inhomogenously broadened i.e.~they have slightly different transition frequencies. For solid-state color centers, the distribution of the emitter frequencies can be modelled as a normal distribution with standard deviation $\Delta\lesssim$ 20 GHz \cite{radulaski2017scalable, banks2018resonant, lukin20194h}. Results of a Monte--Carlo analysis on the transmission and equal-time two-photon correlation through the multi-emitter system are shown in Fig.~\ref{fig:inhom_broadening}(a) for resonant emitters and Fig.~\ref{fig:inhom_broadening}(b) for detuned emitters. We observe an emergence of a large number of very narrow linewidth dips in $g^{(2)}(0; \omega_L)$ which correspond to the subradiant photon blockade that has been studied in CQED systems with two non-identical emitters \cite{radulaski2017photon}. {The occurrence of these dips is due to subradiant states. These highly entangled states that did not overlap with the cavity mode when the emitters were identical, now do overlap with the cavity mode and hence contribute to light emission from the system.} These blockades reach very low $g^{(2)}(0; \omega_L)$ values even for emitters that individually couple to the cavity only weakly. Moreover, for the resonant system, the distribution of the frequencies of the blockade dips ($\omega_B$) reveal that the spread in the frequencies of the subradiant photon blockade is of the order of the inhomogeneous broadening in the emitter frequencies, whereas the frequencies of polaritonic photon blockade are significantly more robust to inhomogeneous broadening in the emitter frequencies albeit with a much larger value of $g^{(2)}(0; \omega_B)$. A similar trend is observed in the detuned system [Fig.~\ref{fig:inhom_broadening}(b)], with the polaritonic dip being robust to inhomogeneous broadening, and the interference-based dips {(identified as the first dip which smoothly plateaus to 1 as $|\omega_L|\to\infty$)} are much more sensitive to the inhomogeneous broadening while reaching very low $g^{(2)}(0; \omega_B)$ values ($\sim0 - 0.1$) similar to the identical-emitter system.

{Finally, our study has uncovered two fundamental tradeoffs in multi-emitter CQED systems which can help inform future experiments and their suitability for quantum information processing applications. Firstly, for a given emitter-cavity coupling strength and cavity decay rate, there exists a tradeoff between the achievable transmission and the depth of photon blockade [measured as $g^{(2)}(0; \omega_B)
$]. Increasing either the cavity-emitter detuning or the number of emitters increases the depth of photon blockade, but also reduces transmission at the blockade frequency. Secondly, a tradeoff exists between the depth of achievable blockade and robustness of the blockade frequency to inhomogeneous broadening --- polaritonic photon blockade, which typically has $g^{(2)}(0; \omega_B) \sim 1$, is robust to inhomogenous broadening, while detuning the emitters from the cavity resonance can allow the multi-emitter CQED system to exhibit the interference-based photon blockade with significantly lower $g^{(2)}(0; \omega_L)$. However, relying on destructive interference of two-photon emissions from various two-excitation eigenstates makes the blockade sensitive to the emitter frequencies. Moreover, the subradiant dips in the photon blockade also provide very low $g^{(2)}(0; \omega_B)$, but the blockade frequencies $\omega_B$ are difficult to engineer without precise control over the emitter frequencies.}

\begin{acknowledgements}
\emph{Acknowledgements.} We thank Shuo Sun and Daniil Lukin for fruitful discussions. We gratefully acknowledge financial support from the Air Force Office of Scientific Research (AFOSR) MURI Center for Attojoule Optoelectronics Award No. FA9550-17-1-0002. RT acknowledges support from Kailath Stanford Graduate Fellowship. MR acknowledges support from Nano- and Quantum Science and Engineering Postdoctoral Fellowship at Stanford University.
\end{acknowledgements}

%

\end{document}


\title{Supplementary information: Photon Blockade in weakly-driven CQED systems with many emitters}

\author{Rahul Trivedi}\email{rtrivedi@stanford.edu}
\affiliation{E. L. Ginzton Laboratory, Stanford University, Stanford CA 94305, USA}
\author{Marina Radulaski}
\affiliation{E. L. Ginzton Laboratory, Stanford University, Stanford CA 94305, USA}
\author{Kevin A. Fischer}
\affiliation{E. L. Ginzton Laboratory, Stanford University, Stanford CA 94305, USA}
\author{Shanhui Fan}
\affiliation{E. L. Ginzton Laboratory, Stanford University, Stanford CA 94305, USA}
\author{Jelena Vu\v{c}kovi\'c}
\affiliation{E. L. Ginzton Laboratory, Stanford University, Stanford CA 94305, USA}

\date{\today}

\maketitle


\section{Transmission statistics for arbitrary bosonic scattering problems}
In this section, we show that for any scattering problem with coherent drive, the transmissitivity and two-particle correlation through the system can be expressed in terms of the single- and two-particle scattering matrices. Henceforth, we will consider the particles as photons without loss of generality. Photonic transport through multi-emitter systems like the ones considered in the main text are a special case of this general problem.

The main issue addressed in this section is the fact that a continuous wave coherent state input is not normalizable, with the photon number in the coherent state being infinitely large even at weak driving amplitudes. Intuitively, analyzing the response of a system to such a state should require computation of scattering matrices with an arbitrary number of input photons. Here, we show that despite this issue of normalizability of the coherent state, a few-photon approximation of the continuous wave coherent state will still give the correct result for the transmission and the two-photon correlation in the limit of a weak coherent drive.

We consider the input state $\ket{\psi_\text{in}}$ to be a pulsed-coherent state given by:
\begin{align}
    \ket{\psi_\text{in}} = \exp\bigg(-\frac{1}{2} \int_{t=-\infty}^{\infty} |\beta(t)|^2 \textrm{d}t \bigg) \bigg[\ket{\text{vac}} + \sum_{k = 1}^\infty \frac{1}{k !} \int_{t_1, t_2 \dots t_k=-\infty}^\infty \prod_{i=1}^k \textrm{d}t_i \beta(t_i)  b_{t_i}^\dagger  \ket{\text{vac}} \bigg]
\end{align}
where $\beta(t) = \beta_0 \text{rect}(t / \tau) \exp(-\textrm{i}\omega_L t)$, with $\text{rect}(t) = 1 \ \textrm{if} \ |t| \leq 1 \ \textrm{and} \ 0 \ \textrm{otherwise}$. In the limit of $\tau \to \infty$, this state approaches the continuous wave coherent state at frequency $\omega_L$. The output state is then given by:
\begin{align}
    \ket{\psi_\text{out}(\omega_L, \tau)} = \exp(-|\beta_0|^2 \tau)\bigg[\ket{\text{vac}} + \sum_{k = 1}^\infty \frac{\beta_0^k}{k!}\ket{\psi_k(\omega_L, \tau)}\bigg]
\end{align}
where
\begin{align}\label{eq:scat_mat_k_ph}
    \ket{\psi_k(\omega_L, \tau)} &= \frac{\beta_0^{-k}}{k!}\sum_{\mu_1, \mu_2 \dots \mu_k}\int_{t_1, t_2 \dots t_k=-\infty}^\infty \int_{t_1', t_2' \dots t_k'=-\infty}^\infty S_{\mu_1, \mu_2 \dots \mu_k}(t_1, t_2 \dots t_k; t_1', t_2' \dots t_k') \prod_{i = 1}^k \beta(t_i') (\mu_i)^\dagger_{t_i} \textrm{d}t_i \textrm{d}t_i' \nonumber\\
    &= \frac{1}{k!} \sum_{\mu_1, \mu_2 \dots \mu_k}\int_{t_1, t_2 \dots t_k=-\infty}^\infty \int_{t_1', t_2' \dots t_k' = -\tau}^{\tau} S_{\mu_1, \mu_2 \dots \mu_k}(t_1, t_2 \dots t_k; t_1', t_2' \dots t_k') \prod_{i=1}^k (\mu_i)^\dagger_{t_i} \exp(-\textrm{i}\omega_L t_i') \textrm{d}t_i  \textrm{d}t_i'
\end{align}
where $\mu_i$ denotes a waveguide or loss channel (i.e. $\mu_i \in \{b, c, l^{(1)}, l^{(2)} \dots l^{(k)}\}$), and $S_{\mu_1, \mu_2 \dots \mu_k}(t_1, t_2 \dots t_k; t_1', t_2' \dots t_k')$ is the scattering matrix element which captures the scattering of $k$ photons in the input waveguide to $k$ photons in the ports $\mu_1, \mu_2 \dots \mu_k$. We note that in the limit of $\tau \to \infty$, the integral with respect to $t_1', t_2' \dots t_k'$ in Eq.~\ref{eq:scat_mat_k_ph} converges. To see this, note that this integral can equivalently be expressed in terms of the frequency-domain scattering matrix:
\begin{align}\label{eq:freq_dom_sca_mat}
    &\lim_{\tau \to \infty}\int_{t_1', t_2' \dots t_k' = -\tau}^{\tau} S_{\mu_1, \mu_2 \dots \mu_k}(t_1, t_2 \dots t_k; t_1', t_2' \dots t_k') \exp[-\textrm{i}\omega_L(t_1' + t_2' \dots t_k')] \textrm{d}t_1'\textrm{d}t_2' \dots \textrm{d}t_k' \nonumber \\&= \int_{\omega_1, \omega_2 \dots \omega_k = -\infty}^\infty {S}_{\mu_1, \mu_2 \dots \mu_k}(\omega_1, \omega_2 \dots \omega_k; \omega_L, \omega_L \dots \omega_L) \exp[-\textrm{i}(\omega_1 t_1 + \omega_2 t_2 \dots \omega_k t_k)] \textrm{d}\omega_1 \textrm{d}\omega_2 \dots \textrm{d}\omega_k
\end{align}
As is outlined in \cite{xu2015input2}, the general structure of the $k$-photon scattering matrix has at most $k$ delta-functions that conserve the total frequency of the input and output photons. Therefore, the $k$ integrals over the input-frequencies in Eq.~\ref{eq:freq_dom_sca_mat} remove all the delta functions, resulting in a completely well-defined and finite integrand.

Finally, also note that for time-independent Hamiltonians such as the multi-emitter CQED system considered in the main text, the $k$ photon scattering matrix $S_{\mu_1, \mu_2 \dots \mu_k}(t_1, t_2 \dots t_k; t_1', t_2' \dots t_k')$ depends only on the differences between the time-arguments, and not on their actual values i.e.~
\begin{align}\label{eq:time_inv_scat_mat}
    S_{\mu_1, \mu_2 \dots \mu_k}(t_1, t_2 \dots t_k; t_1', t_2' \dots t_k') \equiv S_{\mu_1, \mu_2 \dots \mu_k}(0, t_2 - t_1 \dots t_k - t_1; t_1'-t_1, t_2'-t_1 \dots t_k'-t_1)
\end{align}

\subsection{Transmittivity with a weak continuous-wave coherent drive}
\noindent Consider now the computation of the transmissivity $T(\omega; t, \tau)$ through the multi-emitter CQED system at amplitude $\beta_0$:
\begin{align}\label{eq:full_transmittivity}
    T(\omega_L; t, \tau, \beta_0) = \frac{\bra{\psi_\text{out}(\omega_L, \tau)} c_t^\dagger c_t \ket{\psi_\text{out}(\omega_L, \tau)}}{|\beta_0|^2} = \frac{\exp(-2|\beta_0|^2 \tau)}{|\beta_0|^2} \sum_{k=1}^\infty \frac{|\beta_0|^{2k}}{(k!)^2} \bra{\psi_k(\omega_L, \tau)} c_t^\dagger c_t \ket{\psi_k(\omega_L, \tau)}
\end{align}
wherein we have divided the photon flux (number of photons per unit time) in the input state in the input waveguide with the photon flux in the output state in the output waveguide at time $t$. Taking the limit of $\beta_0 \to 0$:
\begin{align}
    \lim_{\beta_0 \to 0} T(\omega_L; t, \tau, \beta_0) = \bra{\psi_1(\omega_L, \tau)} c_t^\dagger c_t \ket{\psi_1(\omega_L, \tau)} .
\end{align}
Now, taking the limit of $\tau \to \infty$, we obtain:
\begin{align}
    \lim_{\tau \to \infty} \bigg[\lim_{\beta_0 \to 0}T(\omega_L; t, \tau, \beta_0)\bigg] =  \bigg |\int_{t_1' = -\infty}^{\infty} S_c(t_1; t_1') \exp(-\textrm{i}\omega_L t_1')\textrm{d}t_1' \bigg|^2.
\end{align}
Finally, using the time-invariance of the system (Eq.~\ref{eq:time_inv_scat_mat}), we immediately see that this limit is independent of $t$:
\begin{align}\label{eq:tran_final}
    T(\omega_L) = \lim_{\tau \to \infty} \bigg[\lim_{\beta_0 \to 0}T(\omega_L; t, \tau, \beta_0)\bigg] = \bigg|\int_{t_1' = -\infty}^{\infty} S_c(0; t_1' - t_1) \exp(-\textrm{i}\omega_L t_1')\textrm{d}t_1' \bigg|^2 = \bigg|\int_{\tau' = -\infty}^{\infty} S_c(0; \tau') \exp(-\textrm{i}\omega_L \tau')\textrm{d}\tau' \bigg|^2.
\end{align}

\subsection{Two-photon correlation with a weak continuous-wave drive}
\noindent The two-photon correlation in the output state is defined by:
\begin{align}
    g^{(2)}(t_1, t_2; \omega_L, \beta_0, \tau) &= \frac{\bra{\psi_\text{out}(\omega_L, \tau)}c_{t_1}^\dagger c_{t_2}^\dagger c_{t_1} c_{t_2} \ket{\psi_\text{out}(\omega_L, \tau)}}{\bra{\psi_\text{out}(\omega_L, \tau)} c_{t_1}^\dagger c_{t_1}\ket{\psi_\text{out}(\omega_L, \tau)}\bra{\psi_\text{out}(\omega_L, \tau)} c_{t_2}^\dagger c_{t_2}\ket{\psi_\text{out}(\omega_L, \tau)}} \nonumber \\
    &=\frac{\exp(2|\beta_0|^2\tau)\sum_{k=2}^\infty \frac{|\beta_0|^{2k}}{(k!)^2} \bra{\psi_k(\omega_L, \tau)}c_{t_1}^\dagger c_{t_2}^\dagger c_{t_1}c_{t_2} \ket{\psi_k(\omega_L, \tau)}}{\bigg[\sum_{k=1}^\infty \frac{|\beta_0|^{2k}}{(k!)^2} \bra{\psi_k(\omega_L, \tau)} c_{t_1}^\dagger c_{t_1} \ket{\psi_k(\omega_L, \tau)}\bigg]\bigg[\sum_{k=1}^\infty \frac{|\beta_0|^{2k}}{(k!)^2} \bra{\psi_k(\omega_L, \tau)} c_{t_2}^\dagger c_{t_2} \ket{\psi_k(\omega_L, \tau)}\bigg]}.
\end{align}
Taking the limit of $\beta_0 \to 0$ (corresponding to a weak coherent drive), we obtain:
\begin{align}
    \lim_{\beta_0 \to 0} g^{(2)}(t_1, t_2; \omega_L, \beta_0, \tau) = \frac{\bra{\psi_2(\omega_L, \tau)}c_{t_1}^\dagger c_{t_2}^\dagger c_{t_1}c_{t_2} \ket{\psi_2(\omega_L, \tau)}}{4\bra{\psi_1(\omega_L, \tau)}c_{t_1}^\dagger c_{t_1} \ket{\psi_1(\omega_L, \tau)}\bra{\psi_1(\omega_L, \tau)}c_{t_2}^\dagger c_{t_2} \ket{\psi_1(\omega_L, \tau)}}.
\end{align}
Next, we take the limit of $\tau \to \infty$ (corresponding to a continuous-wave drive) to obtain. As already shown in the previous subsection,
\begin{align}
    \lim_{\tau \to \infty} \bra{\psi_1(\omega_L, \tau)} c_t^\dagger c_t \ket{\psi_1(\omega_L, \tau)} = \bigg|\int_{\tau' = -\infty}^{\infty} S_c(0; \tau') \exp(-\textrm{i}\omega_L \tau')\textrm{d}\tau' \bigg|^2 = T(\omega_L).
\end{align}
Similarly,
\begin{align}
    \lim_{\tau \to \infty} \bra{\psi_2(\omega_L, \tau)} c_{t_1}^\dagger c_{t_2}^\dagger c_{t_1}c_{t_2} \ket{\psi_2(\omega_L, \tau)} &= \bigg | \int_{t_1', t_2' = -\infty}^{\infty} S_{c, c}(t_1, t_2; t_1', t_2')\exp[-\textrm{i}\omega_L (t_1' + t_2')]\textrm{d}t_1' \textrm{d}t_2' \bigg |^2 \nonumber\\
    &= \bigg |\int_{\tau_1', \tau_2' = -\infty}^{\infty} S_{c, c}(0, t_2 - t_1; \tau_1', \tau_2')\exp[-\textrm{i}\omega_L (\tau_1' + \tau_2')]\textrm{d}\tau_1' \textrm{d}\tau_2' \bigg|^2.
\end{align}
wherein we have used the time-invariance of the multi-emitter system in the last step. Therefore, the two-photon correlation, in the continuous-wave limit, depends only on the difference between the time-instants at which it is being computed. The complete expression for the two-photon correlation is given below:
\begin{subequations}
\begin{align}
    g^{(2)}(t_1, t_2; \omega_L) &=\lim_{\tau \to \infty} \lim_{\beta_0 \to 0} g^{(2)}(t_1, t_2; \omega_L, \beta_0, \tau)  \nonumber \\ 
    &= \frac{1}{4T^2(\omega_L)}\bigg |\int_{t_1', t_2' = -\infty}^{\infty} S_{c, c}(t_1, t_2; t_1', t_2')\exp[-\textrm{i}\omega_L (t_1' + t_2')]\textrm{d}t_1' \textrm{d}t_2' \bigg|^2 \label{eq:g2_without_inv}  \\
    &= \frac{1}{4T^2(\omega_L)}\bigg |\int_{\tau_1', \tau_2' = -\infty}^{\infty} S_{c, c}(0, t_2 - t_1; \tau_1', \tau_2')\exp[-\textrm{i}\omega_L (\tau_1' + \tau_2')]\textrm{d}\tau_1' \textrm{d}\tau_2' \bigg|^2.
\end{align}
\end{subequations}
It can be noted that this limit can be taken for computing any arbitrary $g^{(n)}$ in a similar way.

\section{Computation of single- and two-photon transport}\label{sec:scat_mat}
\noindent In this appendix, we outline the computation of the single- and two-photon scattering matrices [$S_c(t_1; t_1')$ and $S_{c,c}(t_1, t_2; t_1', t_2')$] which are required for computing the transmission and two-photon correlations through the multi-emitter CQED system discussed in the main text. Specifically, we show that cost of computing these scattering matrices is dominated by the cost of diagonalizing the effective Hamiltonian, $H_\text{eff}$, given by:
\begin{align}\label{eq:effective_hamiltonian}
    H_\text{eff} = \bigg(\omega_c - \frac{\textrm{i} \kappa}{2} \bigg) a^\dagger a  + \sum_{n=1}^N \bigg(\omega_n - \frac{\textrm{i}\gamma_n}{2}\bigg) \sigma_n^\dagger\sigma_n  + \sum_{n=1}^N g_n (a \sigma_n^\dagger + \sigma_n a^\dagger),
\end{align}
within the single- and two-excitation subspaces of the multi-emitter CQED system.

As is shown in \cite{xu2015input2, trivedi2018few}, these scattering matrices can be computed by computing the expectations of the cavity annihilation and creation operator ($a$ and $a^\dagger$) evolved under the effective Hamiltonian $H_\text{eff}$ of the multi-emitter system (Eq.~\ref{eq:effective_hamiltonian}):
\begin{subequations}\label{eq:scat_mat_green_func}
\begin{align}
    &S_{c}(t_1; t_1') = -\sqrt{\kappa_b \kappa_c} \bra{\textrm{G}} \mathcal{T}\big\{\tilde{a}(t_1) \tilde{a}^\dagger(t_1') \big\}\ket{\textrm{G}} \\
    &S_{c,c}(t_1, t_2; t_1', t_2') = \kappa_b \kappa_c \bra{\textrm{G}} \mathcal{T}\big\{\tilde{a}(t_1) \tilde{a}(t_2) \tilde{a}^\dagger(t_1') \tilde{a}^\dagger(t_2')\big\}\ket{\textrm{G}}\label{eq:two_ph_scat_mat_green_func}
\end{align}
\end{subequations}
where $\ket{\textrm{G}} = \ket{0}\ket{g_1, g_2\dots g_N}$ is the ground state of the multi-emitter system and $\mathcal{T}$ indicates chronological ordering which, for a given set of time indices, orders the operators in decreasing order of the time indices. The operators $\tilde{a}(t)$ and $\tilde{a}^\dagger(t)$ are given by:
\begin{align}
    \begin{bmatrix} \tilde{a}(t) \\ \tilde{a}^\dagger(t) \end{bmatrix} = \exp(\textrm{i}H_\text{eff}t) \begin{bmatrix} a \\ a^\dagger \end{bmatrix} \exp(-\textrm{i} H_\text{eff} t).
\end{align}
To proceed further, we note that the effective hamiltonian $H_\text{eff}$ commutes with and hence conserves the total excitation number operator $n$:
\begin{align}\label{eq:number_operator}
    n = a^\dagger a + \sum_{i=1}^N \sigma_i^\dagger \sigma_i.
\end{align}
A consequence of this commutation is that the effective Hamiltonian can be expressed in a block-diagonal form:
\begin{align}
    H_\text{eff} \equiv \begin{bmatrix}
        \textbf{H}^{(0)}_\text{eff} &       0               &    0   &   0  & \dots \\
        0 & \textbf{H}_\text{eff}^{(1)} &    0   &   0 & \dots   \\
        0 &       0              & \textbf{H}_\text{eff}^{(2)} & 0 & \dots \\
        0 &        0              &   0                    & \textbf{H}_\text{eff}^{(3)} &  \dots  \\  
        \vdots &        \vdots         &   \vdots               &   \vdots               & \ddots
    \end{bmatrix}
\end{align}
where $\textbf{H}^{(i)}_\text{eff}$ is the projection of the effective Hamiltonian on the $i^\text{th}$ excitation subspace i.e.~space of states which are eigenvectors of $n$ with eigenvalue $i$. Moreover, since $g_i$ are real (and can always be made real by appropriately choosing the phase of the excited states $\ket{e_i}$ of the emitters), $\textbf{H}_\text{eff}$ is complex symmetric. It can thus be diagonalized:
\begin{align}
\textbf{H}_\text{eff}^{(i)} = \textbf{U}^{(i)} \mathcal{D}\{\boldsymbol{\lambda}^{(i)}\}\textbf{U}^{(i)\textrm{T}}
\end{align}
where $\mathcal{D}\{\cdot\}$ of a vector $\textbf{v}$ is a diagonal matrix with elements of $\textbf{v}$ on the diagonal, $\boldsymbol{\lambda}^{(i)}$ is a vector of eigenvalues of $\textbf{H}_\text{eff}^{(i)}$, and $\textbf{U}^{(i)}$ is a matrix with the eigenvectors of $\textbf{H}_\text{eff}^{(i)}$ as its columns. Since $\textbf{H}_\text{eff}^{(i)}$ is complex symmetric, $\textbf{U}^{(i)\text{T}}\textbf{U}^{(i)} = \textbf{U}^{(i)}\textbf{U}^{(i)\text{T}} = \textbf{I}$. We will also denote by $\ket{\phi^{(i)}_k}$ the ket corresponding to the $k^\text{th}$ eigenvector of $\textbf{H}_\text{eff}^{(i)}$ and it immediately follows that $\bra{\phi^{(i)}_k} \phi^{(j)}_l\rangle_\text{T} = \delta_{i, j}\delta_{k, l}$, where $\langle \cdot \rangle_\text{T}$ indicates a `transpose' inner product (as opposed to the complex transpose inner product) in between the two states. Finally, note that once $\textbf{U}^{(i)}$ and $\boldsymbol{\lambda}^{(i)}$ have been computed, it is straightforward to compute the exponential of $\textbf{H}_\text{eff}^{(i)}$:
\begin{align}
    \exp(-\textrm{i}\textbf{H}_\text{eff}^{(i)}t\big) = \textbf{U}^{(i)} \mathcal{D}\{\exp(-\textrm{i}\boldsymbol{\lambda}^{(i)}t\big) \}\textbf{U}^{(i)\text{T}}
\end{align}
Finally, since $a$ and $a^\dagger$ respectively reduce and increase the number of photons inside the cavity, and hence the number of excitations, by 1, their matrix representation has the form:
\begin{align}
    a \equiv 
    \begin{bmatrix} 0  &  \textbf{A}_{0,1} &    0      &     0      & \dots \\
                    0  &   0      & \textbf{A}_{1, 2}  &     0      & \dots \\
                    0  &   0      &    0      &  \textbf{A}_{2, 3}  & \dots \\
                    \vdots &  \vdots & \vdots &  \vdots    & \ddots
    \end{bmatrix} \ \text{and} \
    a^\dagger \equiv 
    \begin{bmatrix}
                    0  &  0 &    0       & \dots \\
                    \textbf{A}_{0, 1}^\dagger  &   0      & 0  & \dots \\
                    0  &   \textbf{A}_{1, 2}^\dagger  &    0      &  \dots \\
                    0  &    0               & \textbf{A}_{2, 3}^\dagger & \dots \\
                    \vdots &  \vdots &   \vdots    & \ddots
    \end{bmatrix}
\end{align}
where $\textbf{A}_{i, i+1}$  is the projections of $a$ onto the direct sum of the $i^\text{th}$ and $(i + 1)^\text{th}$ excitation subspace.

\subsection{Computation of the single photon scattering matrix and transmission}
\noindent Noting that $\exp(-\textrm{i}H_\text{eff} t) \ket{\textrm{G}} = \ket{\textrm{G}}$, the single-photon scattering matrix (Eq.~\ref{eq:scat_mat_green_func}a) can be expressed as:
\begin{align}
    S_c(t_1; t_1') &= -\sqrt{\kappa_b \kappa_c} \bra{\textrm{G}} a \exp(-\textrm{i} H_\text{eff}(t_1 - t_1') a^\dagger \ket{\textrm{G}} \theta(t_1 - t_1') \nonumber \\
                   &= -\sqrt{\kappa_b \kappa_c}\big[\textbf{g}^\text{T}\textbf{A}_{0,1}\textbf{U}^{(1)} \mathcal{D}\{\exp(-\textrm{i}\boldsymbol{\lambda}_1 (t_1 - t_1'))\}\textbf{U}^{(1)\text{T}} \textbf{A}_{0, 1}^\dagger\textbf{g}\big] \theta(t_1 - t_1')
\end{align}
where $\textbf{g}$ is a vector corresponding to $\ket{\text{G}}$. Using Eq.~\ref{eq:tran_final}, we obtain the following expression for the transmittivity:
\begin{align}\label{eq:transmission}
    T(\omega_L) = \kappa_b \kappa_c \bigg|\textbf{g}^T\textbf{A}_{0,1}\textbf{U}^{(1)} \mathcal{D} \bigg\{\frac{1}{ \boldsymbol{\lambda}^{(1)} - \omega_L}\bigg\}\textbf{U}^{(1)\text{T}} \textbf{A}_{0, 1}^\dagger\textbf{g} \bigg|^2 = \kappa_b \kappa_c \bigg|\sum_{i = 1}^{\mathcal{N}_1}\frac{(\bra{\text{G}} a \ket{\phi^{(1)}_i})^2}{\lambda_i^{(1)}-\omega} \bigg|.
\end{align}
where $\mathcal{N}_i$ is the size of the $i^\text{th}$ excitation subspace. We note that this expression for transmission has an intuitively expected form --- the $\omega_L - \lambda_i^{(1)}$ term introduces resonances at the frequencies that match the real part of $\lambda_i^{(1)}$, with the strength of the resonances depending on the matrix element of the cavity annihilation operator $a$ between the corresponding eigenstate $\ket{\phi_i^{(1)}}$ and the ground state $\ket{\text{G}}$.

\subsection{Computation of the two-photon scattering matrix and two-photon correlation}
\noindent Note from Eq.~\ref{eq:two_ph_scat_mat_green_func} that the two-photon scattering matrix is symmetric with respect to an exchange of the time indices $t_1$ and $t_2$, and $t_1'$ and $t_2'$. Therefore, without loss of generality, we will assume $t_1 \geq t_2$ and $t_1' \geq t_2'$. The two-photon scattering matrix then reduces to:
\begin{align}\label{eq:scat_mat_no_time_order}
    &S_{c,c}(t_1, t_2; t_1', t_2') \nonumber \\ &=
    -\kappa_b \kappa_c\begin{cases}
     \bra{\text{G}} a \exp(-\textrm{i}H_\text{eff}(t_1 - t_2)) a \exp(-\textrm{i}H_\text{eff}(t_2 - t_1')) a^\dagger \exp(-\textrm{i}H_\text{eff}(t_1' - t_2')) a^\dagger\ket{\text{G}} & \text{if } t_1 \geq t_2 \geq t_1' \geq t_2' \\ \bra{\text{G}} a\exp(-\textrm{i}H_\text{eff} (t_1 - t_1')) a^\dagger
     \exp(-\textrm{i}H_\text{eff}(t_1' - t_2))a
     \exp(-\textrm{i}H_\text{eff}(t_2 - t_2'))a^\dagger\ket{\text{G}} & \text{if } t_1 \geq t_1' \geq t_2 \geq t_2' \\
     0 & \text{otherwise}\end{cases}
     \end{align}
For $t_1 \geq t_2 \geq t_1' \geq t_2'$, this simplifies to:
\begin{align}\label{eq:explicit_two_ph_scat_mat_2}
    S_{c, c}(t_1, t_2; t_1', t_2') = -\kappa_b \kappa_c  \big[\textbf{g}^\text{T}\textbf{A}_{0,1} \textbf{U}^{(1)}\mathcal{D}\{\exp(-\textrm{i} \boldsymbol{\lambda}^{(1)}(t_1 - t_2))\}\textbf{U}^{(1)\text{T}}\textbf{A}_{1,2}\textbf{U}^{(2)} \mathcal{D}\{\exp(-\textrm{i} \boldsymbol{\lambda}^{(2)}(t_2 - t_1'))\}\textbf{U}^{(2)\text{T}}\times \nonumber \\
    \textbf{A}_{1,2}^\dagger\textbf{U}^{(1)} \mathcal{D}\{\exp(-\textrm{i} \boldsymbol{\lambda}^{(1)}(t_1' - t_2'))\} \textbf{U}^{(1)\text{T}}\textbf{A}_{0,1}^\dagger\textbf{g}\big]
\end{align}
and for $t_1 \geq t_1' \geq t_2 \geq t_2'$, it simplifies to:
\begin{align}\label{eq:explicit_two_ph_scat_mat_1}
 S_{c, c}(t_1, t_2; t_1', t_2') = -\kappa_b \kappa_c  \big[\textbf{g}^\text{T}\textbf{A}_{0,1} \textbf{U}^{(1)}\mathcal{D}\{\text{exp}(-\textrm{i} \boldsymbol{\lambda}^{(1)}(t_1 - t_1'))\}\textbf{U}^{(1)\text{T}}\textbf{A}_{0,1}^\dagger \textbf{A}_{0,1}\textbf{U}^{(1)} \mathcal{D}\{\text{exp}(-\textrm{i} \boldsymbol{\lambda}^{(1)}(t_2 - t_2'))\}\textbf{U}^{(1)\text{T}} \textbf{A}_{0,1}^\dagger\textbf{g}\big]
\end{align}
To compute $g^{(2)}(t_1, t_2; \omega_L)$ as given by in Eq.~\ref{eq:g2_without_inv}, we need to evaluate the integral:
\begin{align}
    \int_{t_1', t_2' = -\infty}^{\infty} S_{c, c}(t_1, t_2; t_1', t_2')\exp[-\textrm{i}\omega_L (t_1' + t_2')]\textrm{d}t_1' \textrm{d}t_2'
\end{align}
Since the two-photon scattering matrix is symmetric with respect to an exchange of the indices $t_1'$ and $t_2'$, it follows that:
\begin{align}
    &\int_{t_1', t_2' = -\infty}^{\infty} S_{c, c}(t_1, t_2; t_1', t_2')\exp[-\textrm{i}\omega_L (t_1' + t_2')]\textrm{d}t_1' \textrm{d}t_2' = 2     \int_{t_1' = -\infty}^\infty \int_{t_2' = -\infty}^{t_1'} S_{c, c}(t_1, t_2; t_1', t_2')\exp[-\textrm{i}\omega_L (t_1' + t_2')]\textrm{d}t_1' \textrm{d}t_2'\nonumber \\
    &=2  \bigg[\int_{t_1' = -\infty}^{t_2} \int_{t_2' = -\infty}^{t_1'} + \int_{t_1'=t_2}^{t_1} \int_{-\infty}^{t_2} \bigg] S_{c, c}(t_1, t_2; t_1', t_2')\exp[-\textrm{i}\omega_L (t_1' + t_2')]\textrm{d}t_1' \textrm{d}t_2'
\end{align}
wherein in the last step we have used the fact that if $t_1 \geq t_2$ and $t_1' \geq t_2'$, then the two photon scattering matrix $S_{c,c}(t_1, t_2; t_1', t_2')$ is 0 unless $t_2 \geq t_1'$ or $t_1 \geq t_1' \geq t_2 \geq t_2'$ (as shown in Eq.~\ref{eq:scat_mat_no_time_order}). These two integrals can be readily evaluated using Eqs.~\ref{eq:explicit_two_ph_scat_mat_2} and \ref{eq:explicit_two_ph_scat_mat_1} to obtain the following expression for the two-photon correlation function $g^{(2)}(t_1, t_2; \omega_L)$:
\begin{align}\label{eq:two_time_corr}
    g^{(2)}(t_1, t_2; \omega_L) = \frac{\kappa_b \kappa_c}{T^2(\omega_L)}\bigg|\textbf{g}^T(\omega_L) \textbf{1} + (\textbf{f}^T(\omega_L) - \textbf{g}^T(\omega_L))\exp(-\textrm{i} (\boldsymbol{\lambda}_1 - \omega_L)(t_1 - t_2))\bigg|^2
\end{align}
with $\textbf{g}(\omega_L)$ and $\textbf{f}(\omega_L)$ being defined by:
\begin{subequations}
\begin{align}
&\textbf{g}(\omega_L) = d \bigg\{\textbf{U}^{(1)\text{T}}\textbf{A}_{0,1}^\dagger\textbf{g}\textbf{g}^\text{T} \textbf{A}_{0,1}\textbf{U}^{(1)} \mathcal{D}\bigg\{\frac{1}{\boldsymbol{\lambda}^{(1)} - \omega_L} \bigg\}\textbf{U}^{(1)\text{T}}\textbf{A}_{0,1}^\dagger \textbf{A}_{0,1}\textbf{U}^{(1)}\mathcal{D}\bigg\{\frac{1}{\boldsymbol{\lambda}^{(1)} - \omega_L} \bigg\} \bigg\} \\
&\textbf{f}(\omega_L) = d \bigg\{\textbf{U}^{(1)\text{T}}\textbf{A}_{1,2}\textbf{U}^{(2)} \mathcal{D}\bigg\{\frac{1}{\boldsymbol{\lambda}^{(2)} - 2\omega_L} \bigg\}\textbf{U}^{(2)\text{T}} \textbf{A}_{1,2}^\dagger\textbf{U}^{(1)} \mathcal{D}\bigg\{\frac{1}{\boldsymbol{\lambda}^{(1)} - \omega_L} \bigg\}\textbf{U}^{(1)\text{T}} \textbf{A}_{0,1}^\dagger\textbf{g}\textbf{g}^\text{T} \textbf{A}_{0,1}\textbf{U}^{(1)} \bigg\}.
\end{align}
\end{subequations}
where $d\{\cdot\}$ of a square matrix $\textbf{A}$ is a vector with the diagonal elements of the matrix. {We note from Eq.~\ref{eq:two_time_corr} that the oscillations in and decay of $g^{(2)}(t_1, t_2; \omega_L)$ with the time-difference $|t_1 - t_2|$ is governed by the real and imaginary part of the eigenvalues of $\textbf{H}_\text{eff}^{(1)}$ (i.e.~eigenvalues of the effective Hamiltonian within the first excitation subspace), and the oscillation amplitude is governed by the detuning between the energies of the single- and two-photon components of input coherent state ($\omega_L$ and $2\omega_L$) with respect to the real part of eigenvalues of the effective Hamiltonian within the single- and two-excitation subspaces. Of particular interest is the equal-time two-photon correlation $g^{(2)}(0; \omega_L) \equiv g^{(2)}(t, t; \omega_L)$, which is given by:
\begin{align}
    g^{(2)}(0; \omega_L) = \frac{\kappa_b \kappa_c}{T^2(\omega_L)}\big|\textbf{g}^T(\omega_L) \textbf{1} \big|^2 = \bigg|\sum_{i=1}^{\mathcal{N}_2} \Gamma_i(\omega_L) \bigg|^2
\end{align}
where $\mathcal{N}_2$ is the size of the two-excitation subspace, and $\Gamma_i(\omega_L)$ is given by:
\begin{align}\label{eq:g20_w}
    \Gamma_i(\omega_L) =  \frac{\sqrt{\kappa_b \kappa_c}}{T(\omega_L)}\Bigg(\frac{\bra{\text{G}}a^2\ket{\phi^{(2)}_i}_\text{T}}{\lambda^{(2)}_i-2\omega_L}\Bigg)\sum_{j=1}^{\mathcal{N}_1}\Bigg(\frac{\bra{\phi^{(2)}_i}a^\dagger\ket{\phi^{(1)}_j}_\text{T}\bra{\phi^{(1)}_j}a\ket{\text{G}}_\text{T}}{\lambda^{(1)}_j - \omega_L}\Bigg)
\end{align}
$\Gamma_i(\omega_L)$ can be interpreted as the contribution of $\ket{\phi^{(2)}_i}$ to the equal-time two-photon emission (relative to the single-photon emission) from the multi-emitter CQED system. It can clearly be seen from this expression that if the laser frequency is resonant with one of the eigenstates in the two-excitation subspace (i.e.~$2\omega_L \approx \text{Re}(\lambda_i^{(2)})$ for some $i \in \{1, 2, 3\dots \mathcal{N}_2\}$), then its contribution to the two-photon emission becomes strong. Moreover, the strength of this contribution also depends on $\bra{\text{G}}a^2 \ket{\phi_i^{(2)}}$, which can be interpreted as the `two-photon' component of the eigenstate. Finally, we note that $\Gamma_i(\omega_L)$ are in general complex and their relative phases are an important factor that govern the strength of the two-photon emission (relative to the single-photon emission). For example,~as shown in the main text, in detuned multi-emitter CQED systems, there are frequencies where all the individual eigenstates strongly emit two-photons, but the individual emissions interfere with each other to give an overall suppression of two-photon emission relative to single-photon emission.}
\subsection{Validating and benchmarking the scattering matrix calculation}
\begin{figure}[t]
    \centering
    \includegraphics[scale=0.23]{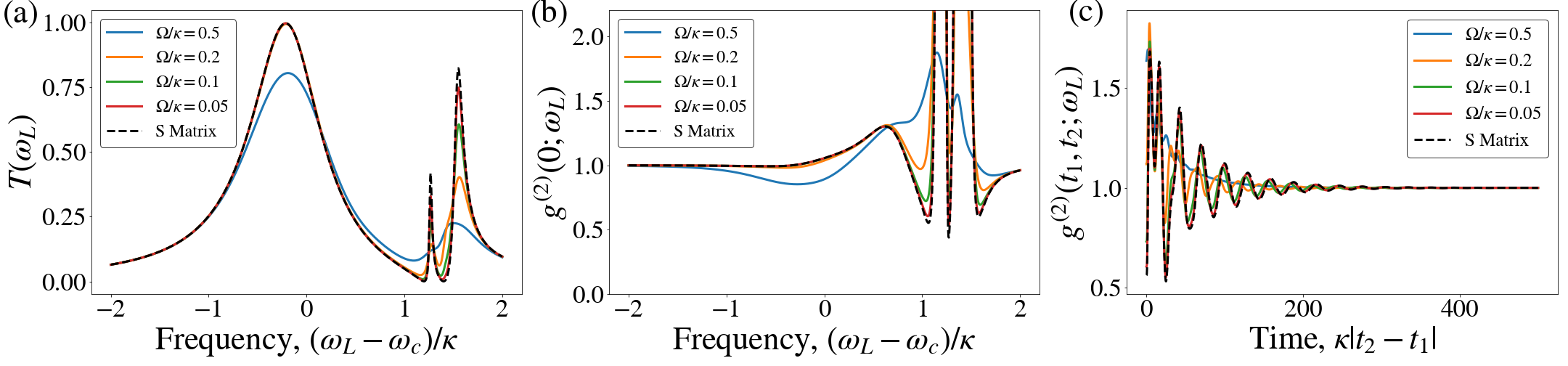}
    \caption{Validation of the scattering matrix calculation against QuTiP for (a) computation of the transmission $T(\omega_L)$, (b) computation of the equal-time two-photon correlation $g^{(2)}(0; \omega_L)$ and (c) computation of the two-photon correlation as a function of the time-difference at $\omega_L -\omega_c = 1.05\kappa $. The validation is done for a two-emitter system with $\kappa = 2\pi \cdot 25$ GHz, $\omega_1 - \omega_c = 2\pi \cdot$ 30 GHz, $\omega_2 - \omega_c = 2\pi\cdot$ 35 GHz and $\gamma = 2\pi \cdot0.3$ GHz.}
    \label{fig:validation}
\end{figure}

In this section, we numerically verify that the transmissivity and two-photon-correlation expressions derived in the previous sections match with a master-equation based simulation of the multi-emitter CQED system. Figure~\ref{fig:validation} shows the comparison between master-equation based simulations (done using the open source python library QuTiP \cite{johansson2012qutip}) with the Scattering matrix approach for a two-emitter system. Within the master-equation framework, the coherent drive is incorporated by adding $\Omega (a + a^\dagger)$ to the system Hamiltonian, where $\Omega =  \sqrt{\kappa_b} \beta_0$ is the driving strength. We see that in the limit of $\beta_0 \ (\text{or } \Omega)\to 0$, the QuTiP simulations agree perfectly with the scattering matrix based simulations, thereby validating the approach.

\begin{figure}[htpb]
    \centering
    \includegraphics[scale=0.3]{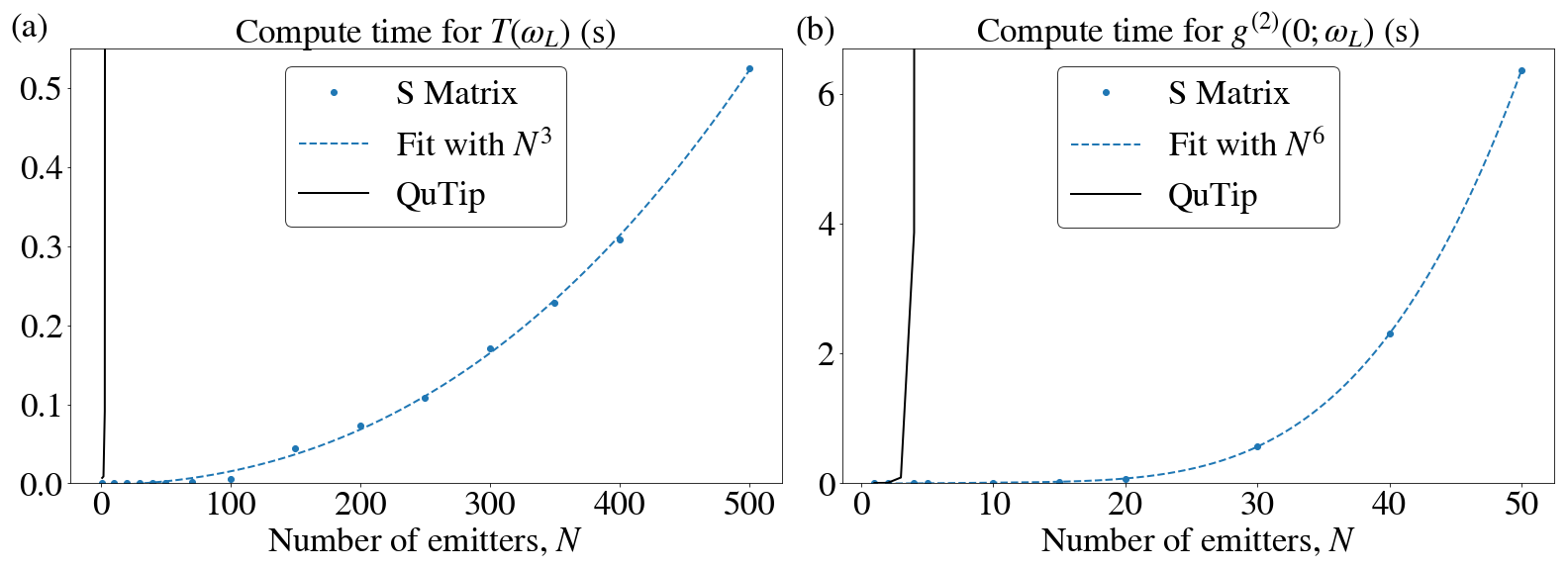}
    \caption{Benchmarking the computation time of the scattering matrix approach with the number of emitters. Note that we benchmark an implementation that can compute photon transport through an arbitrary multi-emitter CQED system (i.e.~it does not exploit the simplification in the eigenvalue computation that can be made when the emitters are identical).}
    \label{fig:benchmark}
\end{figure}

Next, we benchmark the scattering matrix approach --- we compute the time taken to simulate the transmission and two-photon correlation through a system of upto 50 emitters. The results are shown in Fig.~\ref{fig:benchmark} --- we note that even for a system of 50 emitters, computation of the two-photon correlation at a single frequency point takes up to 6s an 2.8 GHz \texttt{Intel Core i7} processor with 16 GB RAM and 8 CPU cores, while utilizing $\sim$1 CPU core. Hence, it would be possible to simulate many more emitters if desired. Moreover, we observe that the compute time for the transmission scales as $N^3$ and the compute time for the two-time correlation scales as $N^6$ --- this is expected theoretically since their computation requires diagonalization of a matrix of size $\sim \mathcal{O}(N)$ and $\sim \mathcal{O}(N^2)$ respectively, and the compute time for diagonalization of a matrix of size $n$ scales as $n^3$.

{
\section{Analysis of multi-emitter CQED systems with identical emitters}

The analysis of the multi-emitter CQED systems can be greatly simplified if all the emitters lie at the same frequencies and couple equally to the cavity mode. In this case, it is possible to reduce the diagonalization of $\textbf{H}_\text{eff}^{(1)}$ and $\textbf{H}_\text{eff}^{(2)}$ to the diagonalization of $3\times3$ and $2\times2$ matrices. In this section, we describe how this diagonalization can be performed, and we present a study of the resulting eigenstates and eigenvalues with respect to the system parameters. Using the outlined method, we also derive analytical results for $T(\omega_L)$ and $g^{(2)}(0; \omega_L)$ in the limit of $N \to \infty$.

\subsection{Diagonalizing $\textbf{H}_\text{eff}^{(1)}$ and $\textbf{H}_\text{eff}^{(2)}$ for identical emitters}
When all the emitters are identical ($\omega_i = \omega_e$, $g_i = g$ and $\gamma_i = \gamma$ for all $i \in \{1, 2 \dots N\}$), the effective Hamiltonian $H_\text{eff}$ (Eq.~\ref{eq:effective_hamiltonian}) can be expressed as \cite{tavis1968exact, tavis1969approximate}:
\begin{align}
    H_\text{eff} = \bigg(\omega_c - \frac{\textrm{i}\kappa}{2}\bigg) a^\dagger a + \bigg(\omega_e - \frac{\textrm{i}\gamma}{2}\bigg) S_z + g (S^\dagger a + a^\dagger S) + \frac{N}{2}\bigg(\omega_e - \frac{\textrm{i}\gamma}{2}\bigg)
\end{align}
where $S_z$ and $S$ are the collective spin operators for all the emitters taken together:
\begin{align}
    S_z = \sum_{i=1}^N \sigma_i^\dagger \sigma_i -\frac{N}{2}, \ S = \sum_{i=1}^N \sigma_i
\end{align}
The Hilbert space of the $N$ two-level systems can be expressed as a direct sum of hilbert spaces of particles with spins $N / 2$, $N / 2 - 1$, $N / 2 - 2 \dots$ using the Clebsh-Gordon expansion \cite{dicke1954coherence}:
\begin{align}
    \bigotimes_{i=1}^N \mathcal{H}_2 = \bigoplus_{i=1}^{\lfloor N/2 \rfloor} \bigoplus_{j=1}^{\mathcal{D}_i} \mathcal{H}_{N + 1 -2i}^{(j)}
\end{align}
where $\mathcal{H}_i^{(j)}$ denotes the Hilbert space of a system with spin $(i - 1) /2$ (i.e.~a system with $i$ states), and $
\mathcal{D}_i$ are the number of such spins required in the expansion:
\begin{align}
    \mathcal{D}_i = \bigg(\frac{N + 1 - 2i}{N + 1}\bigg)\prescript{N+1}{}{C}_{i}
\end{align}
We will denote the basis for $\mathcal{H}_i^{(j)}$, assumed to be the standard spin states, with $\ket{l , m; j}$ where $l = (i - 1)/ 2$ and $m \in \{-l, -l+1, \dots, l-1, l\}$. Note that these states are eigenstates of $S_z$:
\begin{align}
    S_z \ket{l, m; j} = m \ket{l, m;j}
\end{align}
Moreover, the action of the operators $S$ and $S^\dagger$ on this state is given by:
\begin{align}
    &S\ket{l, m; j} = [(l + m)(l - m + 1)]^{1/2} \ket{l, m - 1; j} \\
    &S^\dagger \ket{l, m; j} = [(l - m)(l + m + 1)]^{1/2}\ket{l, m + 1; j}
\end{align}
Note also that the excitation operator $n$ defined in Eq.~\ref{eq:number_operator} can be rewritten as:
\begin{align}\label{eq:number_operator_identical}
    n = a^\dagger a + \frac{S_z}{2} + \frac{N}{2}
\end{align}
and consequently the effective hamiltonian conserves $a^\dagger a + S_z / 2$.\\

\noindent\emph{Single excitation subspace}: Consider now computing the eigenstates of $H_\text{eff}$ within the single excitation subspace --- it follows from Eq.~\ref{eq:number_operator_identical} that this state can only be a superposition of spin states with $m = -N/2$ with 1 photon in the cavity and $m = -(N - 2)/2$ and 0 photons in the cavity:
\begin{align}
    \ket{\phi^{(1)}} = A \ket{1}\ket*{\frac{N}{2}, -\frac{N}{2}; 1} + B\ket{0}\ket*{\frac{N}{2}, -\frac{N-2}{2}; 1} + \sum_{j=1}^{N-1}B_j \ket{0}\ket*{\frac{N-2}{2}, -\frac{N-2}{2}; j}
\end{align}
Substituting this ansatz into the eigenvalue equation $H_\text{eff}\ket{\phi^{(1)}} = \lambda^{(1)}\ket{\phi^{(1)}}$ we immediately conclude that the states of the form $\ket{0}\ket{(N-2)/2, -(N-2)/2; j}$ for $j \in \{1, 2, 3 \dots N-1\}$ are eigenstates with eigenvalue $\lambda^{(1)} = (\omega_e - \textrm{i}\gamma /2 )$ --- these correspond to the `subradiant' states since they do not couple to the cavity mode. The remaining two single-excitation eigenstates are given by $A\ket{1}\ket{N/2, -N/2;0} + B\ket{0}\ket{N/2, -(N-2)/2; 1}$, where $A, B$ and $\lambda^{(1)}$ satisfy the following eigenvalue equation:
\begin{align}\label{eq:single_ex_eigenproblem}
    \begin{bmatrix}
        \omega_c - \textrm{i}\kappa/2 & g\sqrt{N} \\
        g\sqrt{N} & \omega_e - \textrm{i}\gamma / 2
    \end{bmatrix}
    \begin{bmatrix}
    A \\
    B
    \end{bmatrix} = \lambda^{(1)}
    \begin{bmatrix}
    A \\
    B
    \end{bmatrix}
\end{align}
where, as described in section \ref{sec:scat_mat}, we impose the normalization $A^2 + B^2 = 1$. \\

\noindent \emph{Two excitation subspace}: In this case, since $n = 2$, the eigenstates of $H_\text{eff}$ can only be a superposition of states with $m = -N/2$ and 2 photons in the cavity, $m = -(N - 2)/2$ and 1 photon in the cavity and $m = -(N - 4) / 2$ and 0 photons in the cavity.
\begin{align}
    \ket{\phi^{(2)}} = A \ket{2}\ket*{\frac{N}{2}, -\frac{N}{2}; 1} + B \ket{1}\ket*{\frac{N}{2}, -\frac{N - 2}{2}; 1} + \sum_{j=1}^{N-1} B_j \ket{1}\ket*{\frac{N-2}{2}, -\frac{N-2}{2}; j} + C \ket{0}\ket*{\frac{N}{2}, -\frac{N-4}{2}; 1} + \nonumber\\
    \sum_{j=1}^{N-1}C_j \ket{0}\ket*{\frac{N-2}{2}, -\frac{N-4}{2}; j} + \sum_{j=1}^{N(N-3)/2} E_j \ket{0}\ket*{\frac{N-4}{2}, -\frac{N-4}{2}; j}
\end{align}
Substituting this ansatz into the eigenvalue equation $H_\text{eff}\ket{\phi^{(2)}} = \lambda^{(2)}\ket{\phi^{(2)}}$, we immediately conclude that there are $N(N-3)/2$ two-excitation eigenstates of the form $\ket{0}\ket{(N-4)/2, -(N-4)/2; j}$ for $j \in \{1, 2, 3 \dots N(N-3)/2\}$ with eigenvalue $\lambda^{(1)} = \omega_e - \textrm{i}\gamma /2$. These are the `subradiant' states within the two-excitation subspace since they do not couple to the cavity mode. Moreover, there are $2(N - 1)$ states of the form $B_j \ket{1}\ket{(N - 2) /2, -(N - 2)/2; j} + C_j \ket{0}\ket{(N - 2) / 2, -(N - 4) /2; j}$ (i.e.~with at most 1 photon in the cavity) where
\begin{align}
    \begin{bmatrix}
    \omega_e + \omega_c - \textrm{i}(\kappa + \gamma)/2 & g\sqrt{2(N -2)} \\
    g\sqrt{2(N-2)} & 2\omega_e - \textrm{i}\gamma 
    \end{bmatrix}
    \begin{bmatrix}
    B_j \\
    C_j
    \end{bmatrix} =
    \lambda^{(2)}
    \begin{bmatrix}
    B_j \\
    C_j
    \end{bmatrix}
\end{align}
where $B_j^2 + C_j^2 = 1$. Since these states don't have an overlap with two photons in the, they donot contribute to equal-time two-photon emission (Eq.~\ref{eq:g20_w}). Finally, there are three eigenstates of the form $A\ket{2}\ket{N/2, -N/2; 1} + B\ket{1}\ket{N/2, -(N-2)/2; 1} + C\ket{0}\ket{N/2, -(N - 4)/2; 1}$ where:
\begin{align}\label{eq:two_ex_eigenproblem}
    \begin{bmatrix}
    2\omega_c - \textrm{i}\kappa & g\sqrt{2N} & 0 \\
    g\sqrt{2N} & \omega_e + \omega_c - \textrm{i}(\kappa + \gamma)/2 & g\sqrt{2(N -1)} \\
    0 & g\sqrt{2(N-1)} & 2\omega_e - \textrm{i}\gamma 
    \end{bmatrix}
    \begin{bmatrix}
    A \\
    B \\
    C 
    \end{bmatrix} =
    \lambda^{(2)}
    \begin{bmatrix}
    A \\
    B \\
    C
    \end{bmatrix}
\end{align}
with $A^2 + B^2 + C^2 = 1$. These three states are the only states within the two-excitation subspace that contribute to equal-time two-photon emission (Eq.~\ref{eq:g20_w}).

Fig.~\ref{fig:eigenvals} shows the dependence of the single- and two-excitation eigenvalues and the eigenvectors of $H_\text{eff}$ on the detuning $\omega_e - \omega_c$ between the emitters and the cavities and the number of emitters $N$. Note from the eigenvalue plot that increasing $N$ makes the eigenenergies of the multi-emitter system increasingly harmonic, thereby suppressing photon blockade in this system. Additionally, for $N\geq 3$, there are three two-excitation eigenstates that contribute to equal-time two-photon emission (i.e.~have non-zero overlap with two photons in the cavity) as opposed to $N \leq 2$ where there are only two such states.  When the emitters are on resonance with the cavity ($\omega_e = \omega_c$), the additional eigenstate, labelled by $\ket{\phi_0^{(2)}}$, (plotted with a dashed orange line in Fig.~\ref{fig:eigenvals}) has two photons in the cavity with probability $\sim 0.5$ and a very low probability of finding one photon in the cavity whereas the other eigenstates, labelled by $\ket{\phi_\pm^{(2)}}$, have two photons in the cavity with probability $1/4$ and one photon in the cavity with probability $1/2$. Moreover, the eigenvalue associated with $\ket{\phi_0^{(2)}}$, $\lambda_0^{(2)}$, converges to $\omega_e + \omega_c$ (also refer to Eqs.~\ref{eq:two_ph_eigval_asymp} and \ref{eq:two_ph_eigval_asymp_nu}), while the other two eigenvalues (i.e.~$\lambda_\pm^{(2)}$ that are associated with $\ket{\phi_\pm^{(2)}}$) tend to $\pm 2g\sqrt{N}$.  Additionally, as $N \to \infty$, the variation in the eigenvalues and eigenvectors with detuning is reduced since for detuning to have an appreciable impact on the eigenstates and eigenvalues of $H_\text{eff}$, $\omega_e - \omega_c \sim g\sqrt{N}$. Consequently, in the limit of $N \to \infty$, $\ket{\phi^{(2)}_0}$ has a vanishingly small probability of having one photon in the cavity, while $\ket{\phi^{(2)}_\pm}$ have one photon in the cavity with probability $1/4$ and two photons in the cavity with probability $1/2$.
\begin{figure}
    \centering
    \includegraphics[scale=0.23]{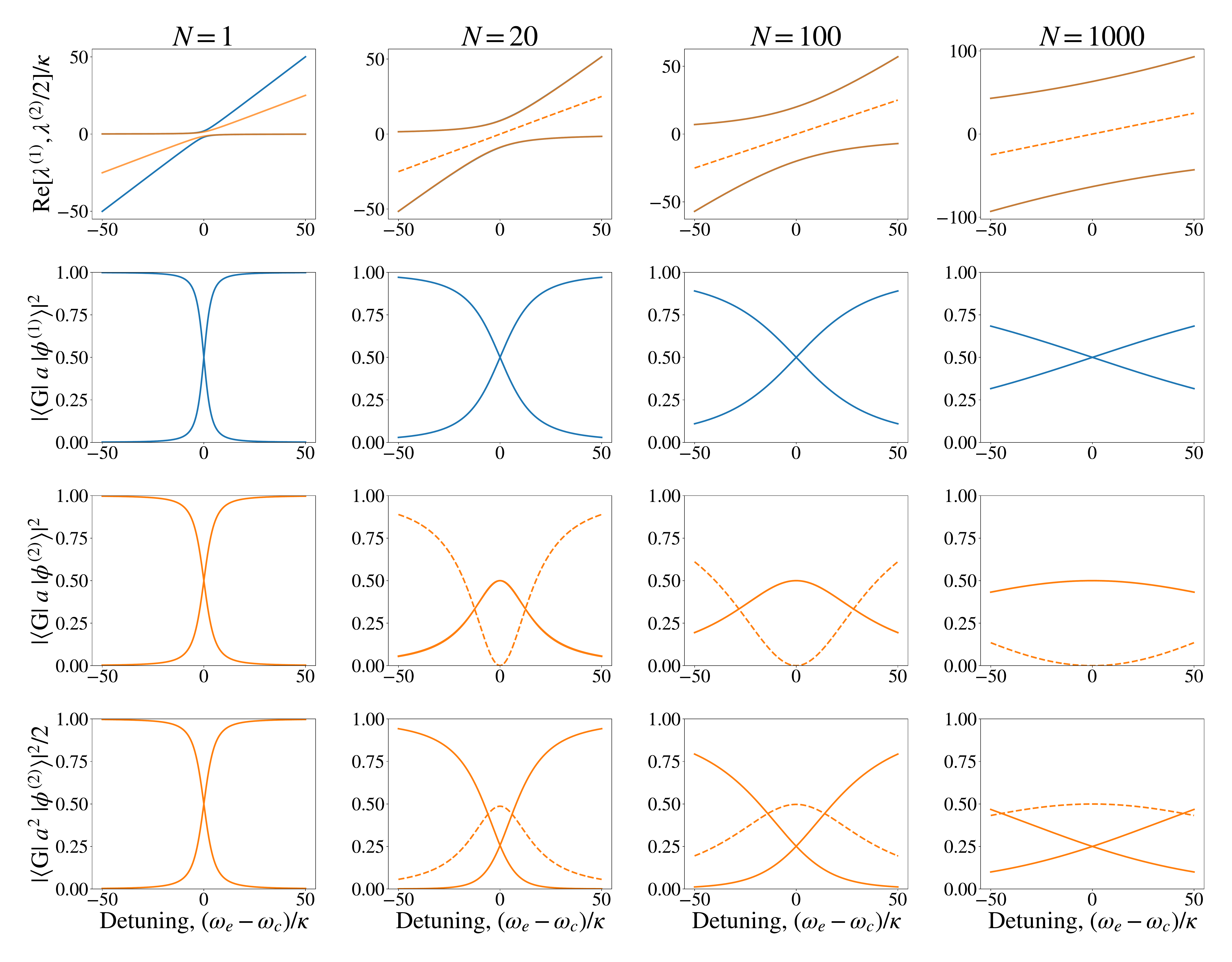}
    \caption{{Single- and two-excitation eigenvalues and eigenvectors of $H_\text{eff}$ as a function of detuning $\omega_e - \omega_c$ and number of emitters $N$. The blue lines indicate that the quantity plotted is associated with the single-excitation subspace, and the orange line indicate that the quantity plotted is within the two-excitation subspace (we only show eigenstates which have a non-zero probability of having two photons in the cavity). For $N \geq 3$, there are three eigenstates in the two-excitation subspace which have a non-zero probability of having two photons in the cavity (and thus contribute to equal-time two photon emission) as opposed to $N \leq 2$ where there only two such states --- the dashed line indicates the quantity plotted is associated with the additional eigenstate that appears when $N \geq 3$. We have assumed $\omega_c = 0$ (which is equivalent to computing the eigenvalues in a frame rotating at $\omega_c$), $g = 2\kappa$ and $\gamma = 0.012\kappa$ in all computations.}}
    \label{fig:eigenvals}
\end{figure}
\subsection{Analytical results for $T(\omega_L)$ and $g^{(2)}(0; \omega_L)$ for $N \to \infty$}
\noindent\emph{Asymptotic forms for the eigenvalues and eigenstates}: Consider first the solution of the eigenvalue problem described in Eq.~\ref{eq:single_ex_eigenproblem} --- it is easily seen by an application of perturbation theory on the matrix in Eq.~\ref{eq:single_ex_eigenproblem} the two eigenvalues $\lambda^{(1)}_+$ and $\lambda^{(1)}_-$ has the following form:
\begin{align}
    \lambda^{(1)}_i = \frac{\omega_e + \omega_c}{2}-\frac{\textrm{i}(\kappa + \gamma)}{4} + \sqrt{N}\mu_i
\end{align}
where
\begin{subequations}
\begin{align}
    &\mu_+ = g + \frac{\Delta^2}{8gN} - \frac{\Delta^4}{128g^3 N^2} + \mathcal{O}(N^{-3}) \\
    &\mu_- = -g - \frac{\Delta^2}{8gN} + \frac{\Delta^4}{128g^3 N^2} + \mathcal{O}(N^{-3})
\end{align}
\end{subequations}
where $\Delta = \omega_e - \omega_c - \textrm{i}(\gamma - \kappa) / 2$ and the corresponding $A_i$ and $B_i$ are given by:
\begin{subequations}
\begin{align}
    A_i = \frac{g\sqrt{N}}{\sqrt{g^2 + (\Delta / 2 + \mu_i\sqrt{N})^2}} \\
    B_i = \frac{\Delta /2 + \mu_i \sqrt{N}}{\sqrt{g^2 + (\Delta / 2 + \mu_i\sqrt{N})^2}}
\end{align}
\end{subequations}
Similarly, for the eigenvalue problem described in Eq.~\ref{eq:two_ex_eigenproblem}, the three eigenvalues $\lambda^{(2)}_+, \lambda^{(2)}_-$ and $\lambda^{(2)}_0$ can be expressed as:
\begin{align}\label{eq:two_ph_eigval_asymp}
    \lambda_i^{(2)} = \omega_e + \omega_c - \frac{\textrm{i}(\kappa + \gamma)}{2}+\sqrt{N}\nu_i
\end{align}
where
\begin{subequations}\label{eq:two_ph_eigval_asymp_nu}
\begin{align}
    &\nu_+ = 2g +\frac{\Delta^2 - 2g^2}{4g N} - \frac{\Delta}{4N\sqrt{N}} + \frac{(\Delta^2 - 2g^2)^2}{256g^3N^2} + \mathcal{O}(N^{-2.5}) \\
    &\nu_- = -2g + \frac{2g^2 - \Delta^2}{4g N}-\frac{\Delta }{4N\sqrt{N}} - \frac{(\Delta^2 - 2g^2)^2}{256g^3N^2} + \mathcal{O}(N^{-2.5})\\
    &\nu_0 = \frac{\Delta}{2N\sqrt{N}} + \mathcal{O}(N^{-2.5})
\end{align}
\end{subequations}
and the corresponding $A_i, B_i$ and $C_i$ are given by:
\begin{subequations}
\begin{align}
    &A_i = \frac{g(2N\nu_i  - \Delta \sqrt{2N})}{\sqrt{(2\nu_i^2 N -\Delta^2)^2+4g^2 N(\Delta^2 + 2N\nu_i^2) - 2g^2(\Delta + \nu_i \sqrt{2N})^2}} \\
    &B_i = \frac{2N\nu_i^2 - \Delta^2}{\sqrt{(2\nu_i^2 N -\Delta^2)^2+4g^2 N(\Delta^2 + 2N\nu_i^2) - 2g^2(\Delta + \nu_i \sqrt{2N})^2}} \\
    &C_i = \frac{g(2N\nu_i  + \Delta\sqrt{2N})}{\sqrt{(2\nu_i^2 N -\Delta^2)^2+4g^2 N(\Delta^2 + 2N\nu_i^2) - 2g^2(\Delta + \nu_i \sqrt{2N})^2}}
\end{align}
\end{subequations}

\begin{figure}
    \centering
    \includegraphics[scale=0.3]{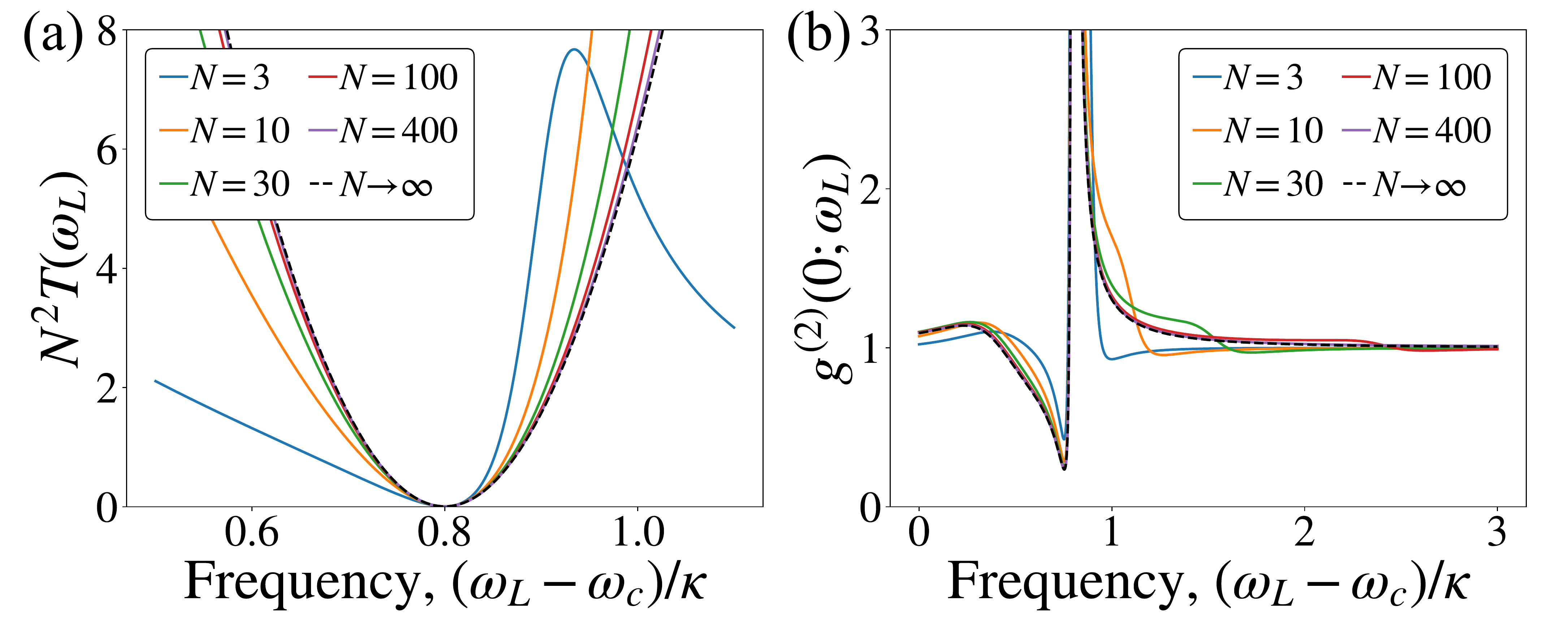}
    \caption{{Validation of the limiting analytical solutions for (a) $T(\omega_L)$ and (b) $g^{(2)}(\omega_L; 0)$ as $N \to \infty$ by comparing them against their numerically computed counterparts for large $N$. $g = 0.2\kappa$, $\omega_e - \omega_c = 0.8\kappa$ and $\gamma = 0.012\kappa$ is assumed in all simulations.}}
    \label{fig:validation_g2_lim}
\end{figure}
\noindent \emph{Asymptotic forms for $T(\omega_L)$ and $g^{(2)}(0; \omega_L)$}: With the above asymptotic forms of the eigenvalues and eigenvectors of $H_\text{eff}$, it is possible to compute the asymptotic forms for the transmission $T(\omega_L)$ and the equal-time two-photon correlation $g^{(2)}(0; \omega_L)$. The transmission $T(\omega_L)$ can be computed using Eq.~\ref{eq:transmission}:
\begin{align}\label{eq:lim_tran}
    \lim_{N\to\infty}T(\omega_L) =\frac{\kappa_b \kappa_c}{g^2 N^2}\bigg|\frac{\omega_L - \lambda_e }{2g} \bigg|^2
\end{align}
where $\lambda_e = \omega_e - \textrm{i}\gamma / 2$. Therefore the transmission at a fixed frequency $\omega$ reduces in proportion to $N^2$ as the number of emitters are increased. Intuitively, this arises due to the polaritonic splitting in the transmission spectrum going to $\infty$ as $N \to \infty$, which consequently results in the transmission being increasingly smaller. Similarly, $g^{(2)}(0; \omega_L)$ can be computed using Eq.~\ref{eq:g20_w} to obtain:
\begin{align}\label{eq:lim_g2}
    \lim_{N \to \infty}g^{(2)}(0; \omega_L) =\bigg |1 - \frac{g^2}{(\omega_L - \lambda_e)(2\omega_L- \lambda_e-\lambda_c)} \bigg |^2
\end{align}
where $\lambda_c = \omega_c - \textrm{i}\kappa / 2$. Fig.~\ref{fig:validation_g2_lim} shows $N^2 T(\omega_L)$ and $g^{(2)}(0; \omega_L)$ for different $N$ alongside with the limiting forms given by Eqs.~\ref{eq:lim_tran} and \ref{eq:lim_g2} --- we obtain excellent agreement between the analytically computed limits and the simulated results for $N^2 T(\omega_L)$ and $g^{(2)}(0; \omega_L)$ for large $N$.

Fig.~\ref{fig:g2_lim_sys_params} shows the impact of detuning between the emitters and the cavity mode, $\omega_e - \omega_c$, and the cavity-emitter coupling constant $g$ on $\lim_{N\to\infty}g^{(2)}(0; \omega_L)$. From Fig.~\ref{fig:g2_lim_sys_params}(a), we see that emission from a resonant multi-emitter system is completely bunched at all frequencies. On increasing the detuning between the emitters and the cavity mode, the interference based antibunching becomes increasingly more pronounced. Also note from Fig.~\ref{fig:g2_lim_sys_params}(b) that increasing the coupling constant between the emitters and the cavity modes necessitates a larger detuning between the emitters and the cavity mode to observe antibunching. This can be intuitively explained by recognizing that the coupling constant $g$ sets the relevant frequency scale for the detuning $\omega_e - \omega_c$ --- consequently, the impact of increasing the coupling strength between the emitters and the cavity mode is similar to the impact of decreasing the detuning between the emitters and the cavity mode.
\begin{figure}[b]
    \centering
    \includegraphics[scale=0.3]{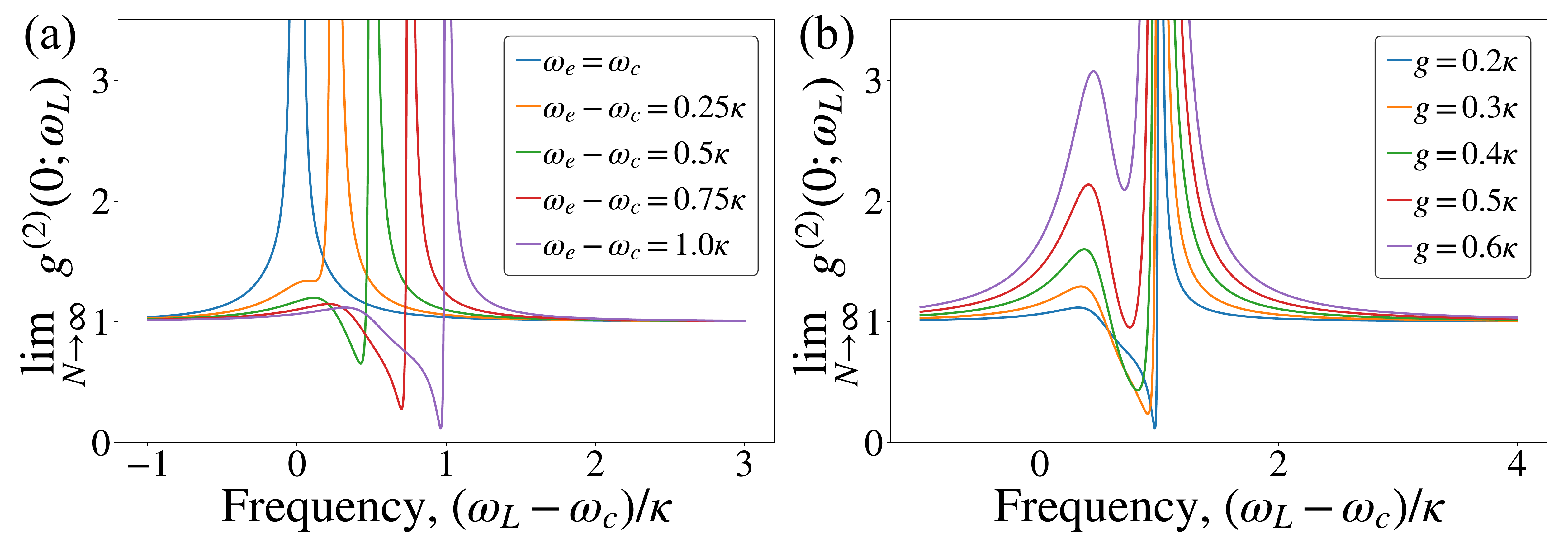}
    \caption{{Impact of (a) detuning $\omega_e -\omega_c$ and (b) coupling strength $g$ between the cavity and emitter on $\lim_{N\to\infty} g^{(2)}(0; \omega_L)$. Note that in (a) we assume $g = 0.2\kappa$ and in (b) we assume $\omega_e - \omega_c = \kappa$. $\gamma = 0.012 \kappa$ is assumed in all computations. }}
    \label{fig:g2_lim_sys_params}
\end{figure}

\section{Study of $g^{(2)}(t_1, t_2; \omega_L)$ as a function of $|t_1 - t_2|$ in detuned multi-emitter CQED systems}
In the main text, we showed that multi-emitter CQED systems exhibited an interference-based photon blockade, where $\text{min}[g^{(2)}(0; \omega_L)]$ improved with $N$, the number of emitters. Fig.~\ref{fig:g2_time}(a) shows the two-photon correlation function $g^{(2)}(t_1, t_2; \omega_L)$ as a function of $|t_1 - t_2|$ at the blockade frequency. It can be seen that the correlation function exhibits oscillation with the time-difference $|t_1 - t_2|$ --- this is different from the time-dependence of a polaritonic blockade which almost monotonically increases to 1 as the time-difference $|t_1 - t_2|$ becomes large. This difference can be attributed to the interference-based nature of the blockade. Moreover, it can also be seen that the settling time of $g^{(2)}(t_1, t_2)$ also decreases with $N$ --- from Fig.~\ref{fig:g2_time}(b), it is evident that it scales as approximately $1 / \sqrt{N}$. This decrease in settling time, as well as its scaling with $N$, can be explained by inspecting Eq.~\ref{eq:two_time_corr}, from which it is easy to see that the correlation time depends on the linewidths of the eigenstates in the single-excitation subspace. Moreover, from Eq.~\ref{eq:single_ex_eigenproblem}, it is easy to show that this linewidth scales as $1/\sqrt{N}$, resulting in the observed decrease in the correlation time.

\begin{figure}[htpb]
    \centering
    \includegraphics[scale=0.27]{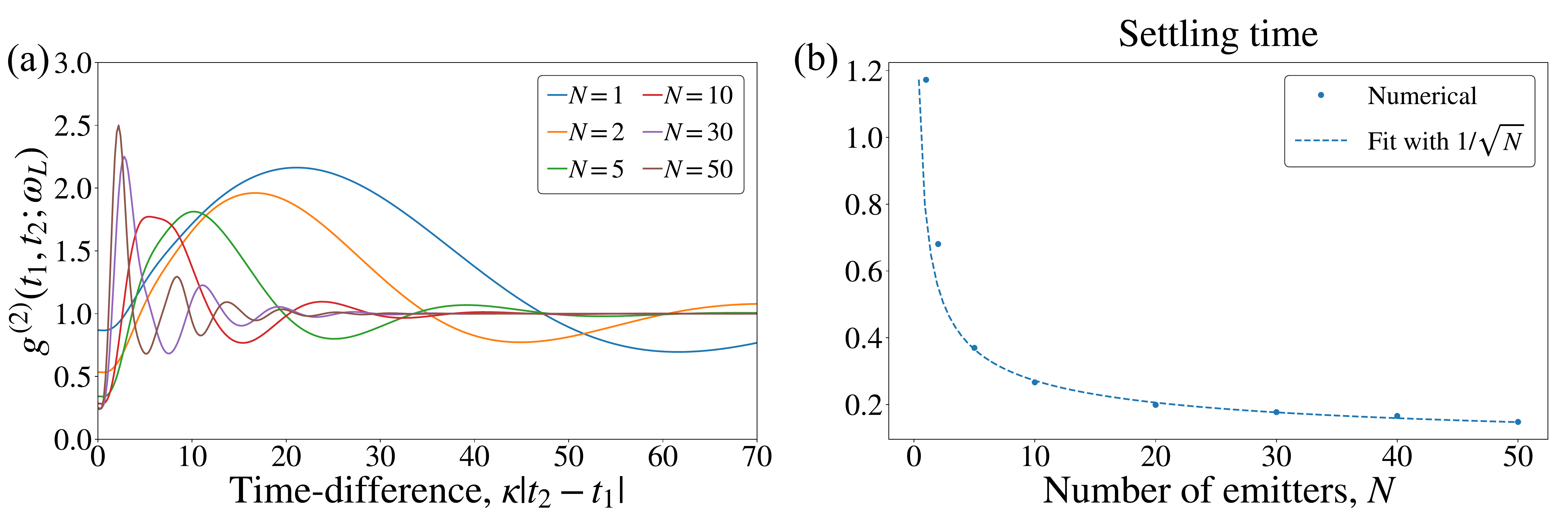}
    \caption{{ (a) $g^{(2)}(t_1, t_2; \omega_L)$ as a function of $|t_1 - t_2|$ at the interference-based blockade frequency. (b) Settling time of $g^{(2)}(t_1, t_2; \omega_L)$ as a function of $N$ as well as its fit with $1 / \sqrt{N}$. $\omega_e - \omega_c = 0.8\kappa$, $g = 0.2\kappa$ and $\gamma = 0.012\kappa$ is assumed in all simulations.}}
    \label{fig:g2_time}
\end{figure}
}